\documentclass[aip,reprint,twocolumn]{revtex4-1}

\pdfoutput=1
\usepackage{graphics}
\usepackage{graphicx}

\usepackage{bm}

\usepackage{hyperref}
\hypersetup{colorlinks=true, citecolor=blue, urlcolor=blue, linkcolor=blue}

\begin{document}

\title{Statistics of small length scale density fluctuations in supercooled viscous liquids}

\author{Ulf R. Pedersen}
\email{ulf@urp.dk}
\affiliation{Glass and Time, IMFUFA, Department of Science and Environment, Roskilde University, Postbox 260, DK-4000 Roskilde, Denmark}

\date{\today}


\begin{abstract}
Many successful theories of liquids near the melting temperature assume that small length scale density fluctuations follow Gaussian statistics. In this paper I present numerical investigations of fluctuations in the supercooled viscous regime using an enhance sampling method. I present results for the single component Lennard-Jones liquid, the Kob-Andersen binary mixture, the Wahnstr{\"o}m binary mixture, the Lewis-Wahnstr{\"o}m model of ortho-terphenyl and for the TIP4P/Ice model of water. Results show that the Gaussian approximation persist to a good degree into the supercooled viscous regime, however, the approximation is less accurate at low temperatures. The analysis suggest that non-Gaussian fluctuations are related to crystalline configurations. Implications to theories of the glass transition are discussed.
\end{abstract}

\maketitle

\section{Introduction}
Small length scale density fluctuations in normal homogeneous liquids above the melting temperature obey Gaussian statistics
over many orders of magnitude.\cite{hum96,cro97} This dependence underlies many successful theories of liquids.\cite{pra77,hum96,cha93,kar07} A Gaussian hypothesis is sometimes assumed for the supercooled regime where the liquid approach a glass transition.\cite{ang95,zac01,deb01,dyr06,dyr06b,edi12,bir13} In this study I directly investigate to what extend Gaussian statistics of small length scale density fluctuations persist into the supercooled viscous regime near the glass-transition. Viscous liquids are highly nontrivial as characterized by the three non's:\cite{dyr06b} non-exponential relaxation of equilibrium fluctuations, non-Arrhenius temperature dependence of structural relaxation time, and nonlinear out-of-equilibrium relaxation. Thus, it is not obvious that Gaussian statistics will persist into the supercooled viscous regime.

The motivation is an ongoing debate about the origin of slow dynamics in structural glass-formers.\cite{ber11,ber05,edi12,sch18} In general, liquids can be cooled below melting temperature due to the existence of a free-energy barrier in the form of a critical nucleus.\cite{bec35} The dynamics of a supercooled liquid near the glass-transition is dramatically slower than near the melting temperature. If dynamics were governed by a fixed free energy barrier the slowdown would follow an Arrhenius law. However, many liquid have super-Arrhenius behavior  (the first ``non'').
Experiments and computer studies of model liquids have shown that dynamics become increasingly spatially heterogeneous upon cooling.\cite{don82,sch91,kob97,wid05} It is an appealing idea that the dynamical heterogeneity is linked to geometric arrangements of locally preferred structures of well-packed particles.
Several studies have identified accumulation of locally preferred structures.\cite{tan98,edi00,wid05,shi06,cos07,roy08,ped10,gal10,pas15,tur18,tur18b} This gives a picture of a less homogeneous structure with non-Gaussian small length scale density fluctuations. A disadvantages of ``locally preferred structure'' approach is that it is system specific. In this paper I propose to study statistics in the collective density field. This is a generic approach that can be applied to widely different systems. This is demonstrated by investigating systems belonging to chemically different classes.

In contrast to the aforementioned structural origin of the glass-transition, some explanations regard structures as less important.\cite{tur61,dyr06b,hed09,cha10} 
The motivation is the experimental observation that the structure factor is similar in the normal- and in the supercooled liquid regime. The quadratic scaling law of the temperature dependency of the relaxation time \cite{gar02,hed09,cha10,rit03} originates from generic kinetic contained models\cite{fre84,rit03}. These models have trivial thermodynamic statistics, but nontrivial slow dynamics leading to a glass-transition. This picture suggests that statistics of small length scale density fluctuations near the glass transition inherit the Gaussian statistics of the normal liquid regime.

In this study I examine the statistics of small length scale density fluctuations for the single component Lennard-Jones (LJ)\cite{lj24} model, the binary Kob-Andersen mixture (KABLJ)\cite{kob94}, the Wahnstr{\"o}m binary mixture (WABLJ)\cite{wah91}, a coarse grained model o-terphenyl (LWoTP)\cite{lew93} and the TIP4P/Ice\cite{aba05} water model. Enhanced sampling molecular dynamics methods are used to sample statistics into the wings of the distributions. Results show that the Gaussian hypothesis is fair in the supercooled regime, however, deviations are more significant at low temperatures. The analysis suggest that non-Gaussian features are related to first-order transitions to crystals.

The remainder of the paper is organized as follows. Section \ref{Formalism} introduce the formalism used the describe density fluctuations and some theory of the Gaussian hypothesis. Section \ref{Method} describe numerical methods used for enhanced sampling of the density field and give descriptions of the investigated models (e.i.\ energy surfaces). Section \ref{Results} present the results, and implications are discussed in Section \ref{Discussion}.

\section{Formalism and theory}\label{Formalism}
In experiments (X-ray or neutron scattering) and theories of the liquid states it is often convenient to work in reciprocal space. In this section I will give the formalism used to describe density fluctuations in both reciprocal $k$-space and a subvolume.

\subsection{The collective density field in $k$-space}
Consider a liquid of $N$ particles located at ${\bf R}=\{{\bf r}_1,{\bf r}_2,\ldots,{\bf r}_N\}$ in a volume $V$ with periodic boundaries so that the thermodynamic density is $\rho=N/V$. Let the real-space density field be $\rho({\bf r})=\sum_n^N\delta({\bf r}_n-{\bf r})$ where $\delta$ is Dirac's delta function. The collective density field in reciprocal space is then defined as the Fourier transform of the real-space density field:
$ \rho_\mathbf{k} = \int_V d\mathbf{r} \rho(\mathbf{r})\exp(-i\mathbf{k}\cdot\mathbf{r})/\sqrt{N} $
where $\mathbf{k}=k\hat{\mathbf{k}}$ is the scattering- or wave vector (sometimes the letter ``$q$'' or ``$Q$'' is used). The $1/\sqrt{N}$ factor ensure system size scale invariance of amorphous configurations (liquids). The scaling is $\sqrt{N}$ for configurations with long-range translational order (crystals) along the $\hat\mathbf{k}$ direction. For a system of $N$ point particles in a periodic orthorhombic cell the field can be written as
\begin{equation}
\rho_\mathbf{k} =\sum_{n=1}^N\exp(-i\mathbf{k}\cdot\mathbf{r}_n)/\sqrt{N}
\end{equation}
where $\mathbf{k}=(2\pi n_x/L_x,2\pi n_y/L_y,2\pi n_z/L_z)$, $n_x$, $n_y$ and $n_z$ are integers, and $L_x$, $L_y$ and $L_z$ are the length of the volume that confines the liquid ($V=L_xL_yL_z$).

Due to the anisotropy of liquids the investigation can be limited to $k$-vectors along the x-direction without loss of information. For a given cubic box with a size of $L=L_x=L_y=L_z$ we consider vectors of lengths $k=2\pi n/L$ where $n$ is an integer ($n_x=n$ and $n_y=n_z=0$). We note that the anisotropy of the liquid is in principle destroyed by the constraint of the {\em an}isotropic periodic boundaries. However, such effects are expected to be small and are ignored in this study.

In the following we consider the probability distribution $P(|\rho_{\bf k}|)$ where $|\rho_{\bf k}|$ is the norm of the collective density field. We do not need to consider the tedious two-dimensional complex plane since the $P(\rho_{\bf k})$ distribution is radial symmetric for a liquid. The second moment $S_\mathbf{k} = \langle |\rho_\mathbf{k}|^2 \rangle$ is the structure factor routinely measured in scattering experiments. If statistics of density fluctuations follow Gaussian (G) statistics then probability distribution of $|\rho_{\bf k}|$ is
\begin{equation}\label{gauss_rhok}
P_G( | \rho_\mathbf{k} |) = 2|\rho_\mathbf{k}| \exp(-| \rho_\mathbf{k} |^2/{S_\mathbf{k}})/S_\mathbf{k}.
\end{equation}
The central limit theorem dictates that in the thermodynamic limit density fluctuations become Gaussian (see also discussion in Section \ref{Discussion}): $P( | \rho_\mathbf{k} |)\rightarrow P_G( | \rho_\mathbf{k} |)$ for $N\rightarrow\infty$. Thus, we limit our analysis to {\emph{small length scale}} fluctuations by studying systems of about 100 to 1000 particles (unless otherwise stated). It have been shown that a small system can represent viscous dynamics of larger system.\cite{heu08}

The fourth moment $\langle|\rho_\mathbf{k}|^4\rangle$ of the Gauss distribution $P_G(|\rho_\mathbf{k}|)$ equals $2\langle|\rho_\mathbf{k}|^2\rangle^2$. Thus, we define a non-Gaussian parameter for the $|\rho_\mathbf{k}|$ fluctuations as
\begin{equation}\label{nonGauss}
\alpha_{\rho_\mathbf{k}} = \langle|\rho_\mathbf{k}|^4\rangle/2\langle|\rho_\mathbf{k}|^2\rangle^2-1.
\end{equation}
This parameter quantifies deviations from Gaussian statistics near the center of the distribution, $|\rho_\mathbf{k}|\simeq0$. Deviations in the tails of the distribution cannot be expected to be represented by this parameter. For this, higher order moments are relevant.

\subsection{The density fluctuations in a subvolume}
The central limit theorem dictates that non-Gaussian feature are more pronounced in smaller systems. Thus, it can be illustrative to investigate density fluctuations in small subvolumes of a larger system.
We define a subvolume though the function $h({\bf r})$ so that $h$ is unity inside the volume and zero outside. The collective density field in this subvolume can then be written as $
\rho^{(h)}_{\bf k} = \sum^N_n \exp(-i{\bf k}\cdot {\bf r}_n)h({\bf r}_n)/\sqrt{V_h}$
where $V_h$ is the size of the subvolume. The $k=0$ relates the actual density $\rho_h=N_h/V_h$ in the subvolume. Here $N_h=\sum_n^N h(r_n)$ is the number of particles in the subvolume. The Gaussian approximation of the $\rho_h$ density fluctuations is
\begin{equation}
P_G(\rho_h)=\exp(-[\rho_h-\langle\rho_h\rangle]^2/2m_2)/\sqrt{2\pi m_2}.
\end{equation}
where $m_2=\langle(\rho_h-\langle\rho_h\rangle)^2\rangle$ is the variance.
For this distribution the fourth central moment $m_4=\langle(\rho_h-\langle\rho_h\rangle )^4\rangle$ equals $3m_2^2$. Thus, we define a non-Gaussian parameter as
\begin{equation}\label{nonGauss_h}
\alpha_{\rho_h} = m_4/3m_2^2-1.
\end{equation}

\begin{figure}
	\begin{center}
		\includegraphics[width=0.99\columnwidth]{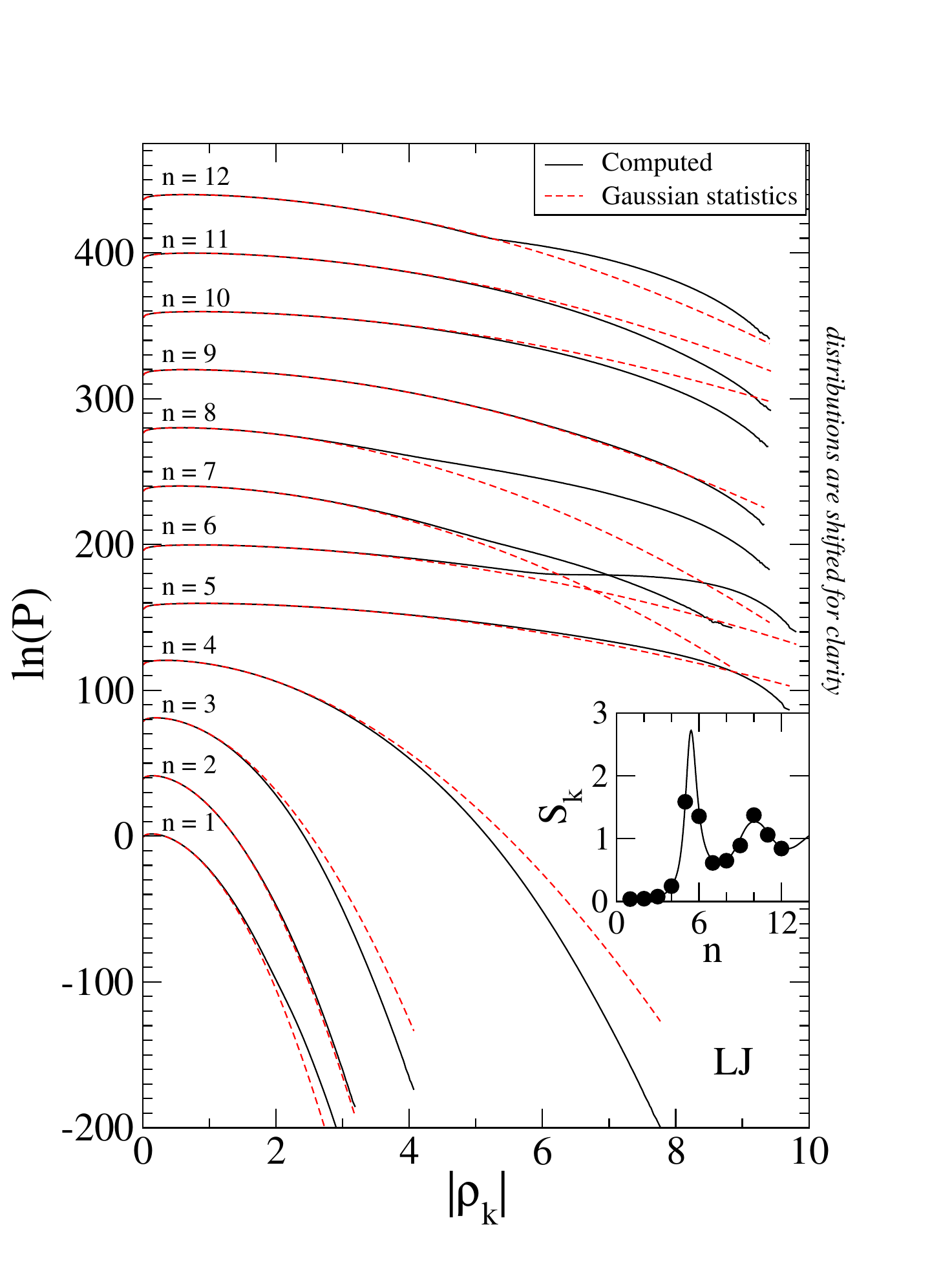}
	\end{center}
	\caption{\label{lj108} Probability distribution $P(|\rho_\mathbf{k}|)$ on a logarithmic scale with $\mathbf{k}=(2\pi n/5.0273,0,0)$ for the single component LJ model in the normal liquid regime. The solid black lines are the distribution function computed from reweighed biased simulations and the red dashed lines are the corresponding Gaussian predictions (Eq.\ \ref{gauss_rhok}). The distributions have been shifted vertically for clarity. The insert show the structure factor where dots indicate the investigated $k$-vectors.}
\end{figure}

\begin{figure}
	\begin{center}
		\includegraphics[width=0.3\columnwidth]{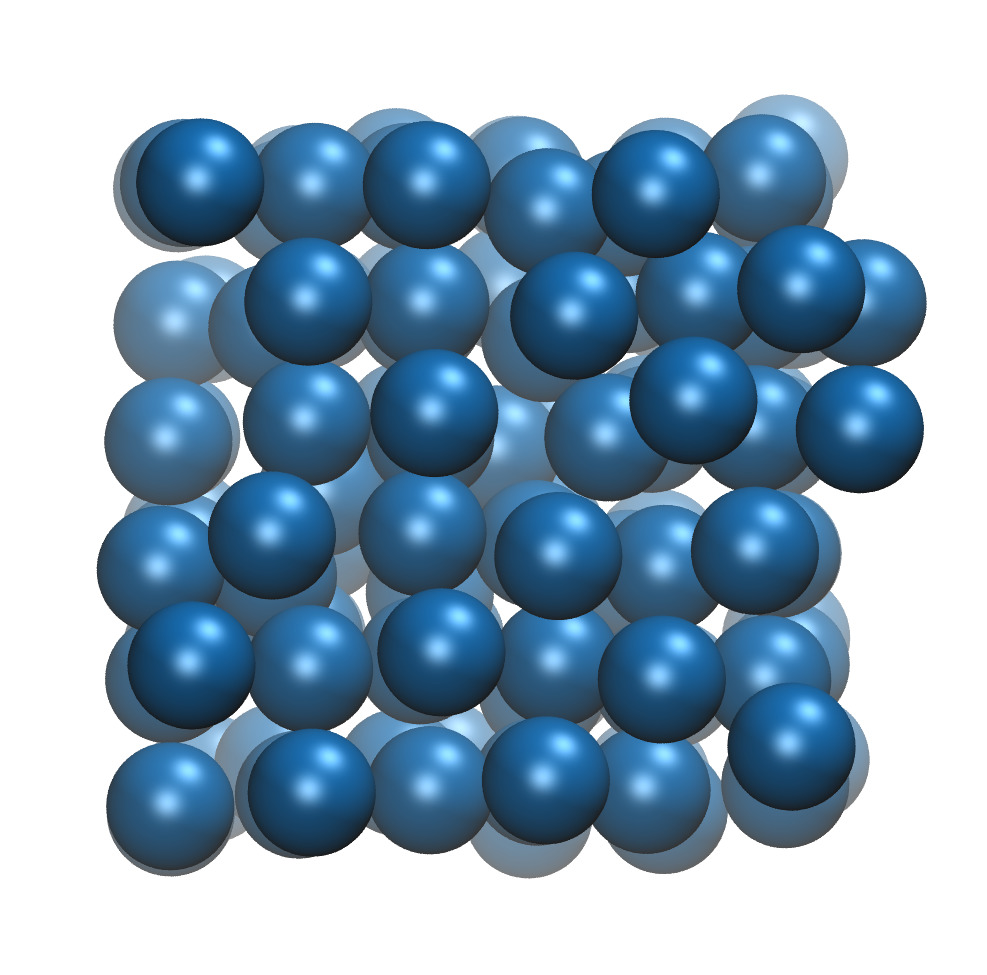} \includegraphics[width=0.2\columnwidth]{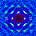} \newline
		\includegraphics[width=0.3\columnwidth]{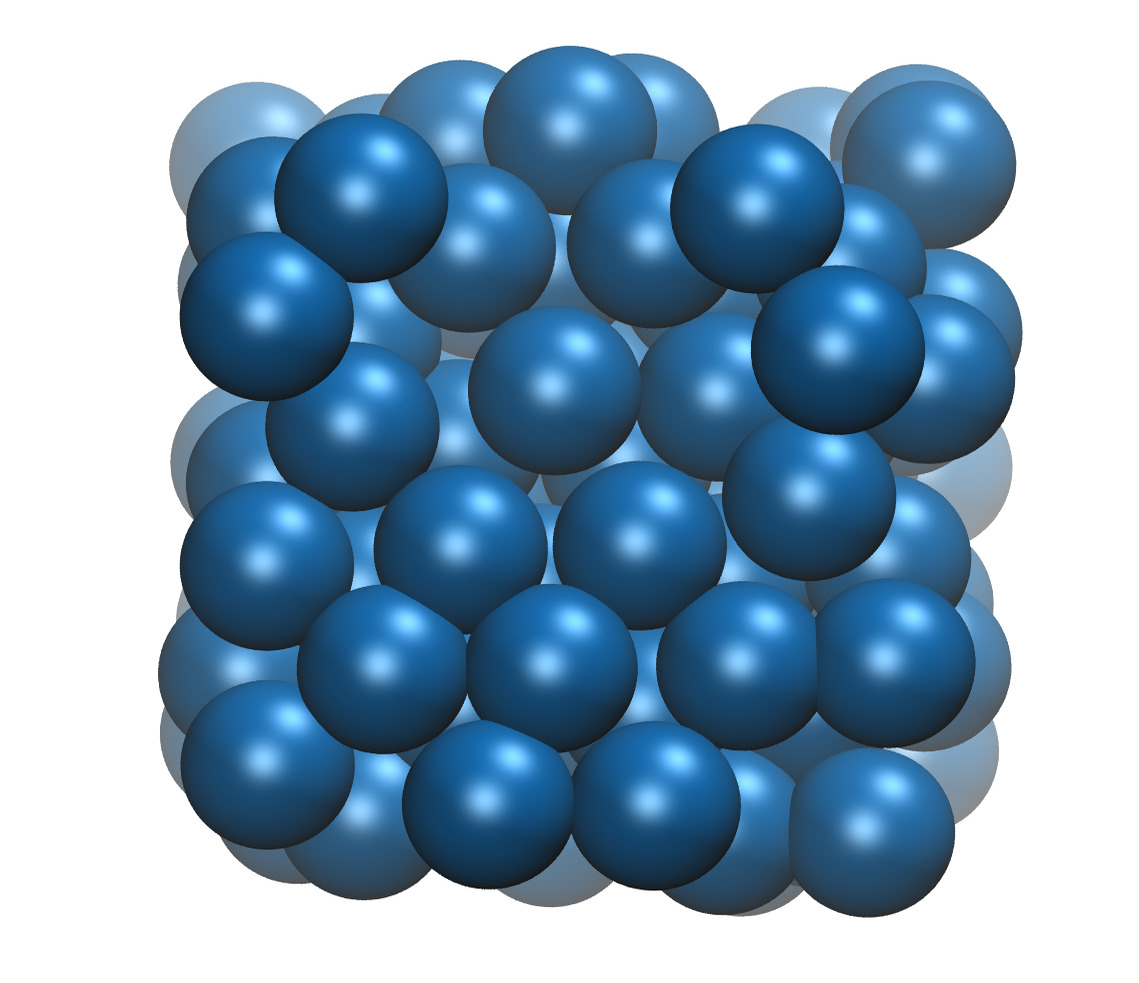} \includegraphics[width=0.2\columnwidth]{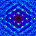}  \newline
		\includegraphics[width=0.3\columnwidth]{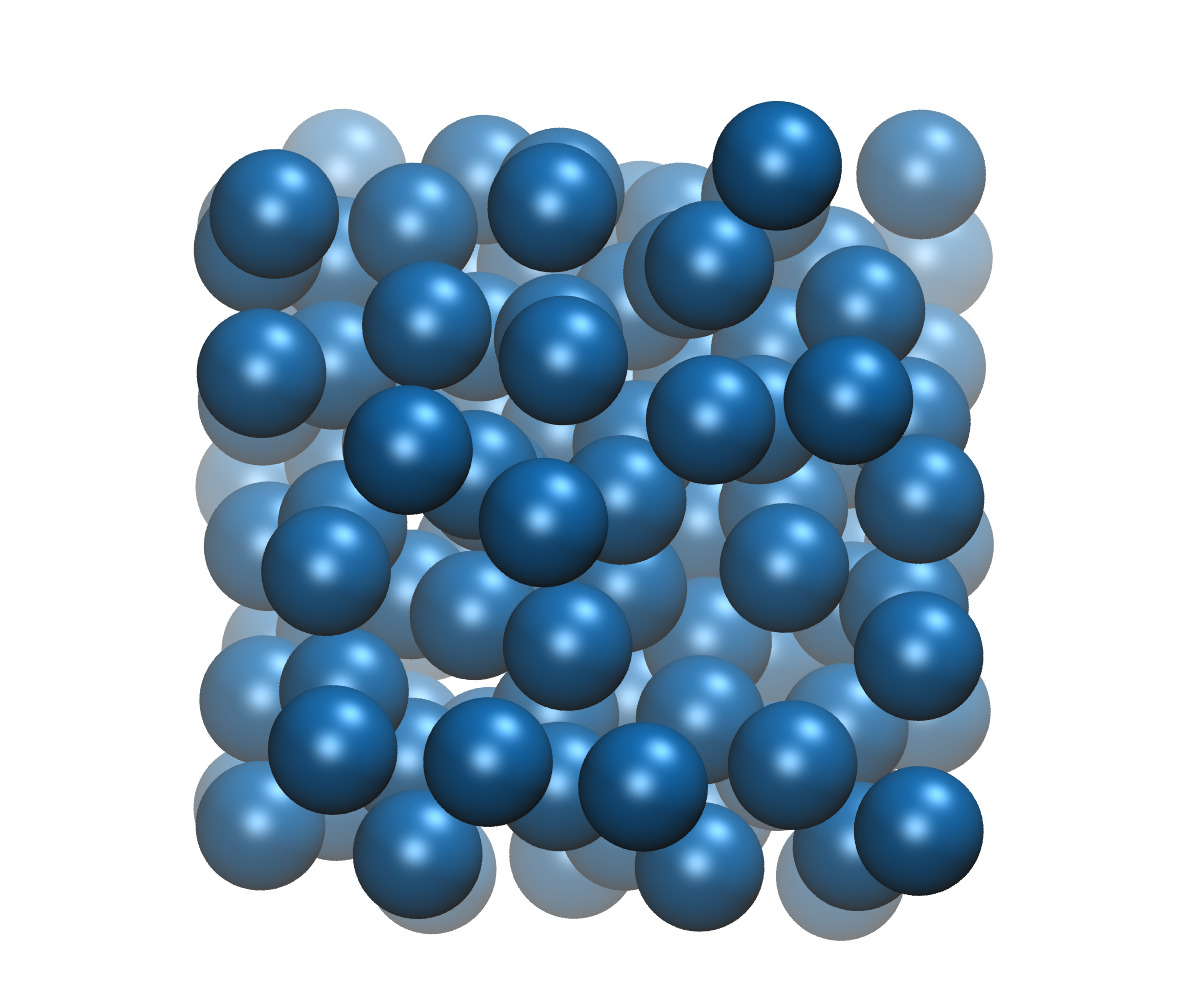} \includegraphics[width=0.2\columnwidth]{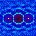}  \newline
		\includegraphics[width=0.3\columnwidth]{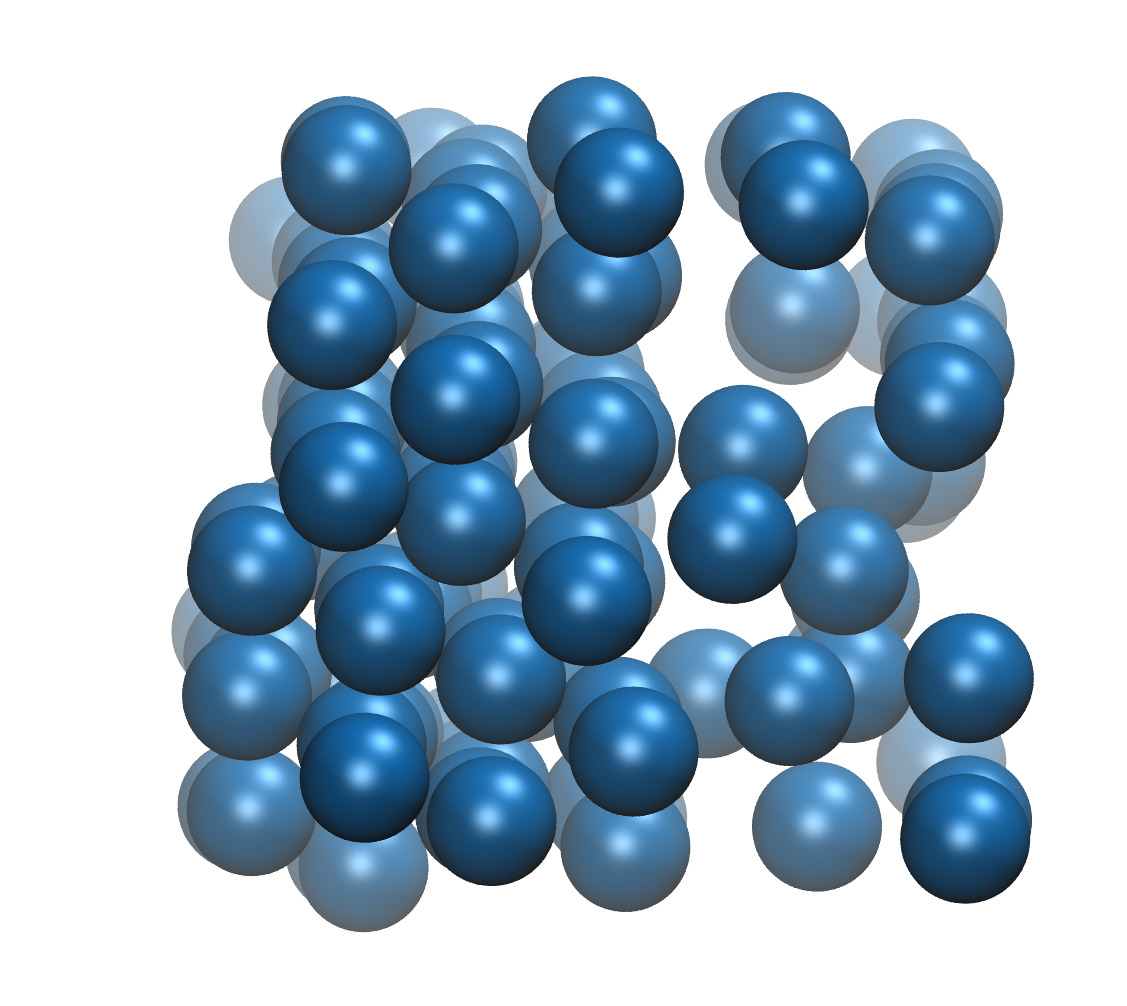} \includegraphics[width=0.2\columnwidth]{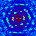} \newline
	\end{center}
	\caption{\label{ljConfs} Configurations taken from the non-Gaussian tails of distributions shown on Fig.\ \ref {lj108}. The left panels show a representative configuration, and the right panels show the scattering spectra $S_\mathbf{k}$ in the $xy$-plane (average over several configurations). From top down the panels shows representations taken from biased simulations with wave vectors with $n=6$, $n=8$, $n=10$ and $n=1$.} 
\end{figure}

\begin{figure}
	\begin{center}
		\includegraphics[width=0.9\columnwidth]{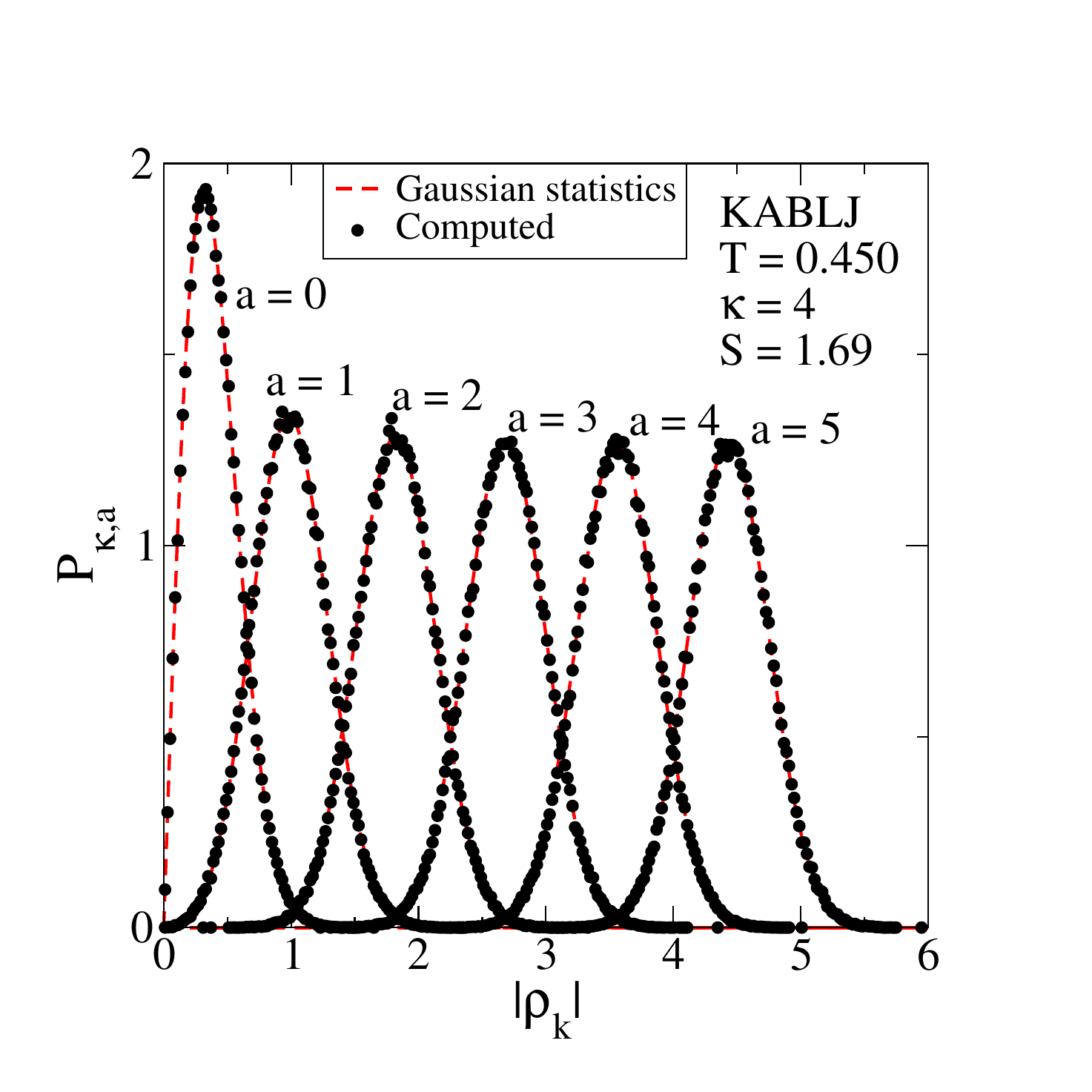}
	\end{center}
	\caption{\label{KABLJ_umb} The distribution function $P_{\kappa a}(|\rho_{\bf k}|)$ in a series of $|\rho_{\bf k}|$ biased simulations ($\kappa=4$ and $a=\{0,2,3,4,5\}$) of the KABLJ liquid in the supercooled regime. The red dashed lines show the Gaussian prediction (Eq.\ \ref{gaussUmbrella}; no fitting parameters). The agreement is good. This kind of data have been reweighed to get the $P(|\rho_{\bf k}|)$ distributions on Fig.\ \ref{rhok_stats}.}
\end{figure}

\section{Methods and models}\label{Method}
To highlight non-Gaussian features, I suggest to apply a potential that bias the system towards rare configurations that would not be sampled otherwise. Specifically, a harmonic potential is added to the Hamiltonian that will bias the system towards large $\rho_{\bf k}$ values. The Gaussian hypothesis can then either be investigated directly on statistics of the biased simulations, or by re-weighting statistics of a series of simulations (referred to as the ``umbrellas sampling method'' \cite{frenkel2002}). Below is a description of the suggested method:

\subsection{Sampling rare fluctuations of the collective density field}
Let ${\mathcal H}({\mathbf R},\dot{\mathbf R})$ be the Hamiltonian of a given system. To sample rare $\rho_\mathbf{k}$ fluctuations at some density and temperature we simulations a Hamiltonian with added harmonic bias field\cite{ped13}:
\begin{equation}
 {\mathcal H}_{\kappa a}({\mathbf R},\dot{\mathbf R}) = {\mathcal H}({\mathbf R},\dot{\mathbf R}) + \kappa [ |\rho_\mathbf{k}({\mathbf R})| - a ]^2/2
\end{equation}
where $\kappa$ is a spring constant and $a$ is an anchor point of the bias field. By reweighing we can get that the $|\rho_{\bf k}|$ probability distribution of the unperturbed system:
\begin{equation}\label{reweight}
 P(|\rho_\mathbf{k}|) = \mathcal{N}_{\kappa a}P_{\kappa a}(|\rho_\mathbf{k}|)\exp(\kappa[|\rho_\mathbf{k}|-a]^2/2k_BT) 
\end{equation}
where $P_{\kappa a}(|\rho_\mathbf{k}|)$ is the distribution of the Hamiltonian with the harmonic bias field. 
For a series of overlapping distributions (with different $a$'s and $\kappa$'s) the normalization constants $\mathcal{N}_{\kappa a}$ can be determined numerically using the iterative multistate Bennett acceptance ratio (MBAR) method.\cite{shirt2008}
Alternatively, the distribution function $P_{\kappa a}$ can be investigated directly. By combining equations \ref{gauss_rhok} and \ref{reweight} we get that the Gaussian approximation predicts
\begin{equation}\label{gaussUmbrella}
 P^{(\kappa a)}_G(|\rho_\mathbf{k}|) = 2|\rho_\mathbf{k} |\exp(-| \rho_\mathbf{k} |^2/S_\mathbf{k}-\kappa'[|\rho_\mathbf{k}|-a]^2)/\mathcal{N}^{(\kappa a)}_G S_\mathbf{k}
\end{equation}
where
\begin{equation}
 \mathcal{N}^{(\kappa a)}_G=\frac{\exp(-\frac{A^2}{S_\mathbf{k}\kappa'})}{1+S_\mathbf{k}\kappa'}\left[\exp(-A^2)+A\pi^\frac{1}{2}[1+\textrm{erf}(A)]\right],
\end{equation}
$A=a\kappa'/\sqrt{S_\mathbf{k}^{-1}+\kappa'}$, $\kappa'=\kappa/2k_BT$ and $\textrm{erf}(x)=2\pi^{-\frac{1}{2}}\int_0^x\exp(-s^2)ds$ is the error function.
The first moment of biased distribution is
\begin{equation}\label{biasMean}
 \langle |\rho_\mathbf{k}|\rangle_{P^{(\kappa a)}_G} = \frac{\exp(-a^2\kappa')}{\mathcal{N}2a\kappa'[1+S_\mathbf{k}\kappa']^2}[2S_\mathbf{k}[a\kappa']^2+B[1+\textrm{erf}(A)]]
\end{equation}
where $B=A\pi^{\frac{1}{2}}e^{A^2}(1+S\kappa'(1+2a^2\kappa'))$. Results of the Gaussian approximation can be used as initial guesses for the iterative MBAR method.\cite{shirt2008} In practice this lead to fewer iterations before reaching convergence.

To perform molecular dynamics simulations forces on particles from the bias field needs to be evaluated. The total force acting on particle $j$ is \cite{ped13}
\begin{equation}\label{force1}
 {\bf F}^{(\kappa a)}_j={\bf F}_j^{(0)}-\kappa ( |\rho_{\bf k}| - a ) \nabla_j |\rho_{\bf k}|
\end{equation}
where ${\bf F}_j^{(0)}$ is the force of the unbiased Hamiltonian, and
\begin{equation}\label{force2}
\nabla_j |\rho_{\bf k}| = -{\bf k}\Re[i\rho_{\bf k}^*   \exp(-i {\bf k}\cdot{\bf r}_j)]/|\rho_{\bf k}|.
\end{equation}
Although the force on particle $j$ depends on the positions of all particles it is possible to design an $N$-scaling algorithm: First loop over all particles to compute $\rho_{k}$ and then loop over all particles again to get particle forces using Eqs.\ \ref{force1} and \ref{force2}. The algorithm can be parallelize to several processes since both the computation of $\rho_{\bf k}$ and the ${\bf F}_j^{(\kappa a)}$ forces involve sums of independent contributions (assuming the same for ${\bf F}_j^{(0)}$).
Computational efficiency of the algorithm is crucial, since we wish to conduct long-time simulations in the viscous regime where dynamics are slow.

\subsection{Sampling rare density fluctuations in a subvolume}
The overall idea of the above mentioned method for computing rare fluctuations of collective density field can be reused to sample rare density fluctuations in a subvolume. In order to perform molecular dynamics with a bias field we define a continuous quantity $\tilde{N}_h$ that is strongly correlated with the number of particles $N_h$ inside the volume $h$. In practice this is done by using a switching functions on the boarders of the volumes as described in Ref.\ \cite{pat10}. The unbiased $P(\rho_h)$ distribution is obtained by reweighing biased $P_\textrm{bias}(\tilde{N}_h)$ distribution using the MBAR method\cite{shirt2008}. For binary mixtures a bias potential can be applied to both kinds of particles. Again statistical information of the unbiased system can then be determined with reweighing.

\begin{figure}
	\begin{center}
		\includegraphics[width=0.45\columnwidth]{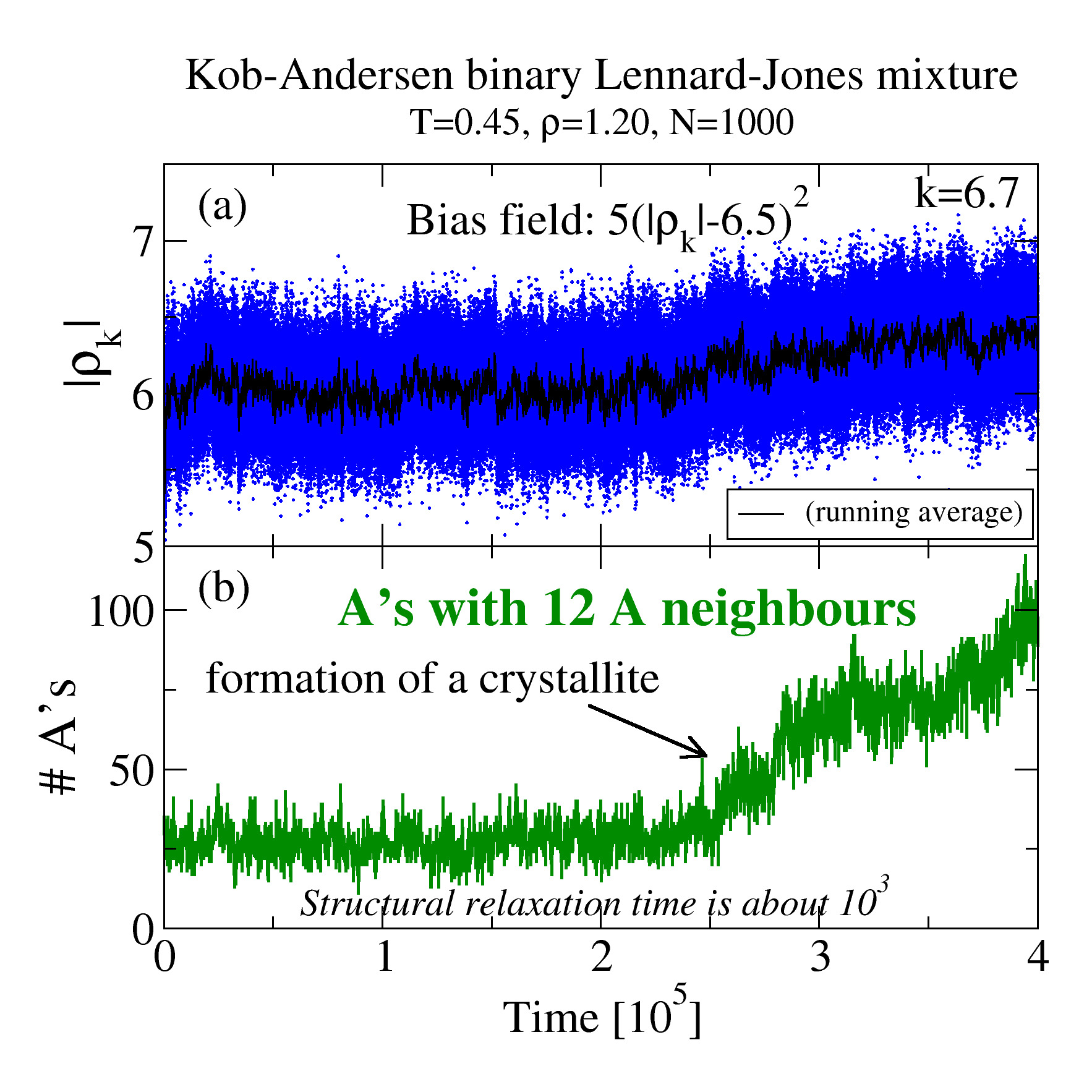}
	    \includegraphics[width=0.45\columnwidth]{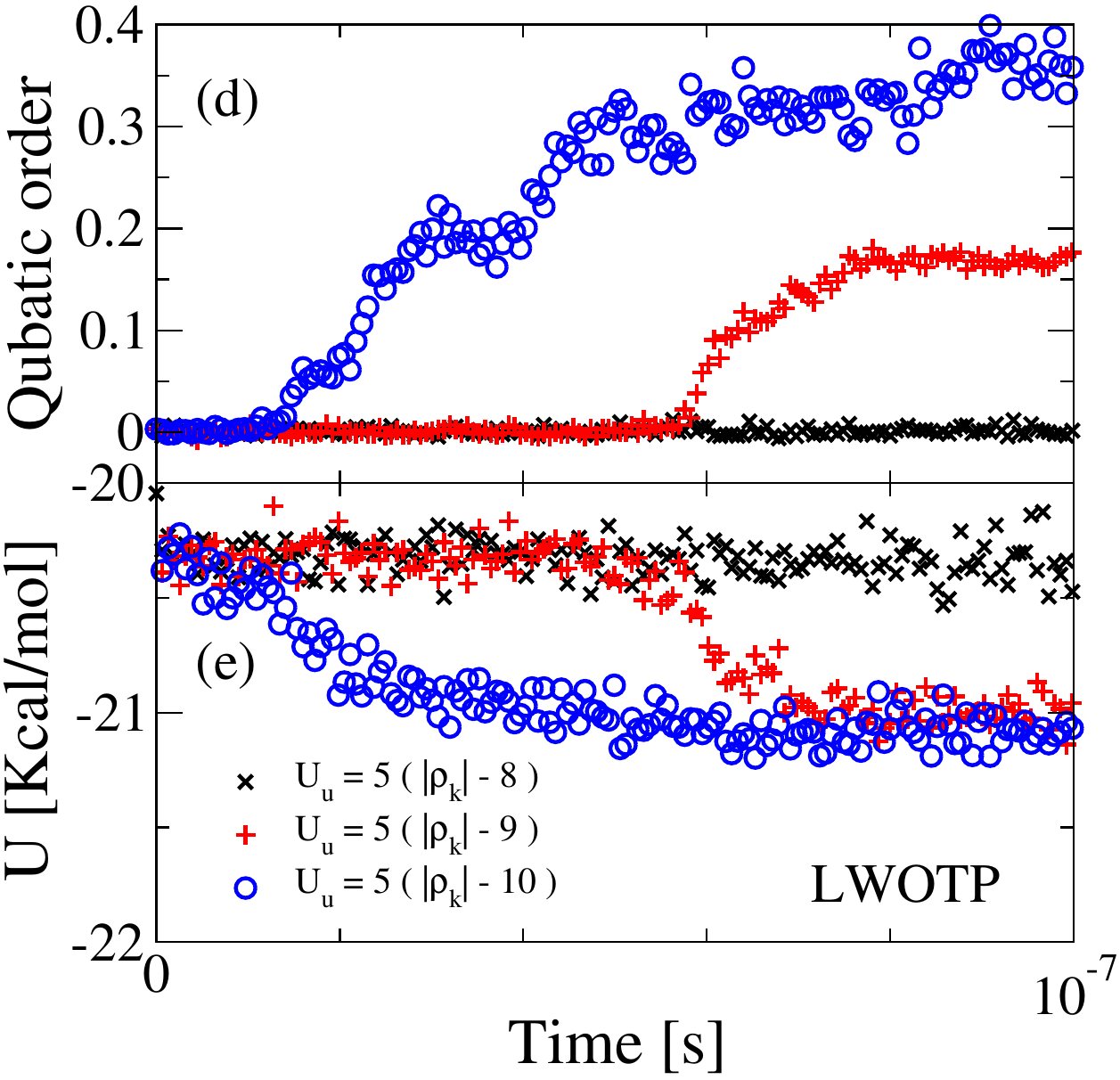} \newline
		\includegraphics[width=0.45\columnwidth]{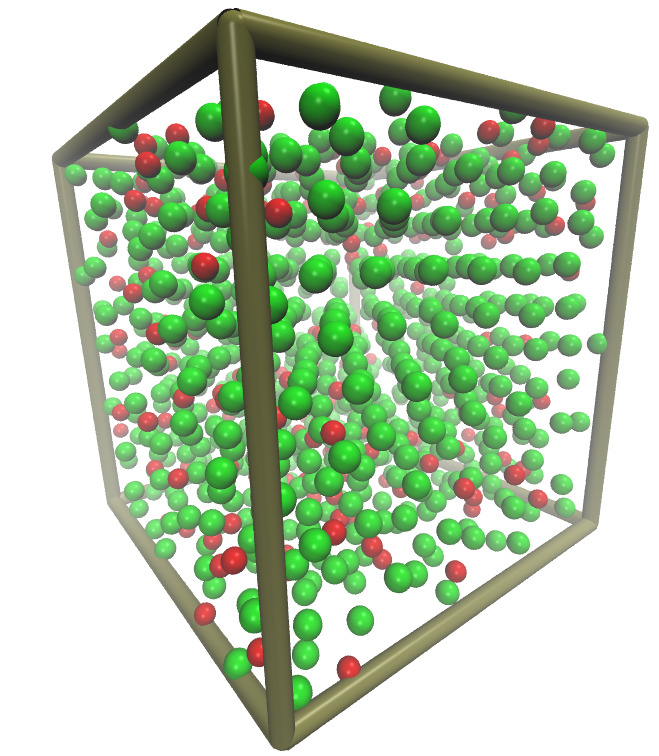}
		\includegraphics[width=0.45\columnwidth]{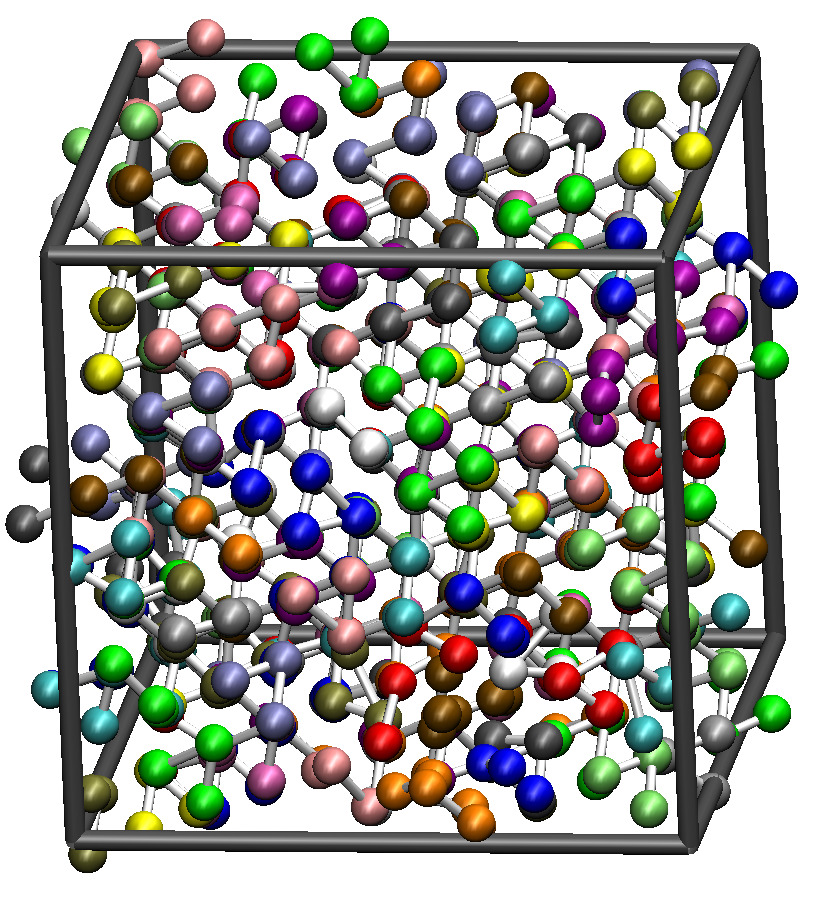}	    
	\end{center}
	\caption{\label{KABLJ_umbCry} Examples of crystallizing trajectories. (a) The $|\rho_{\bf k}|$ trajectory in a simulation with the bias field $5(|\rho_{\bf k}|-6.5)^2$ added to the Hamiltonian of the KABLJ mixture. A crystallite is formed in the last third of the simulation. The crystallite consist of pure A's and crystallization event is accompanied by a phase separation. (b) The number of A's that have 12 A's in the first neighbor shell. This order-parameter is a indicator of the crystallization. (c) The configuration in the last step of the simulation. The larger A's are colored green, while the smaller B's are red. (d) The qubatic order-parameter\cite{ped11b} and (e) the potential energy in biased simulations of the LWoTP model. (f) The final configuration of a trajectory where a crystal is formed. Each molecule have been give an individual color to tell them apart.}
\end{figure}

\begin{figure*}
	\begin{center}
		\includegraphics[width=0.4\textwidth]{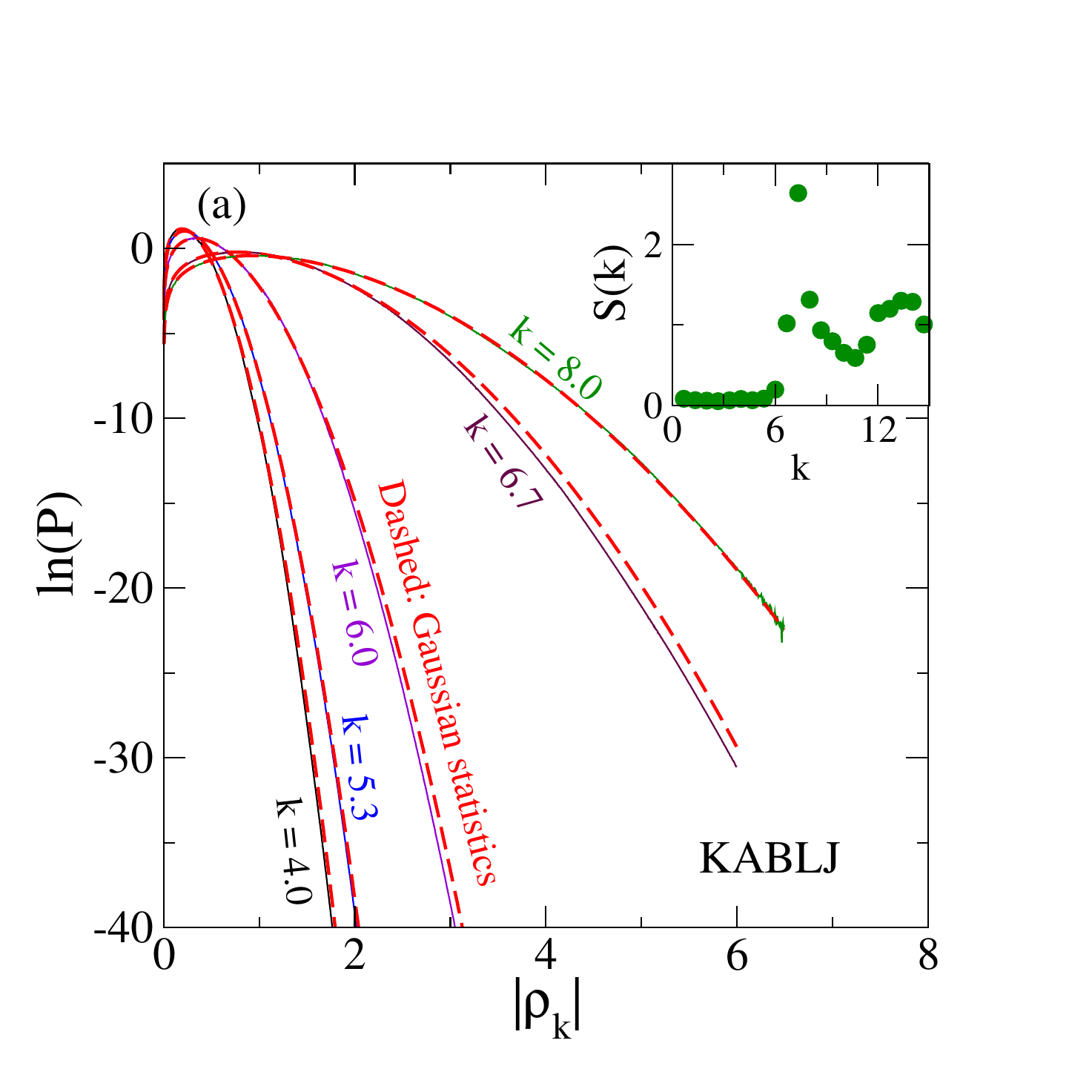}
		\includegraphics[width=0.4\textwidth]{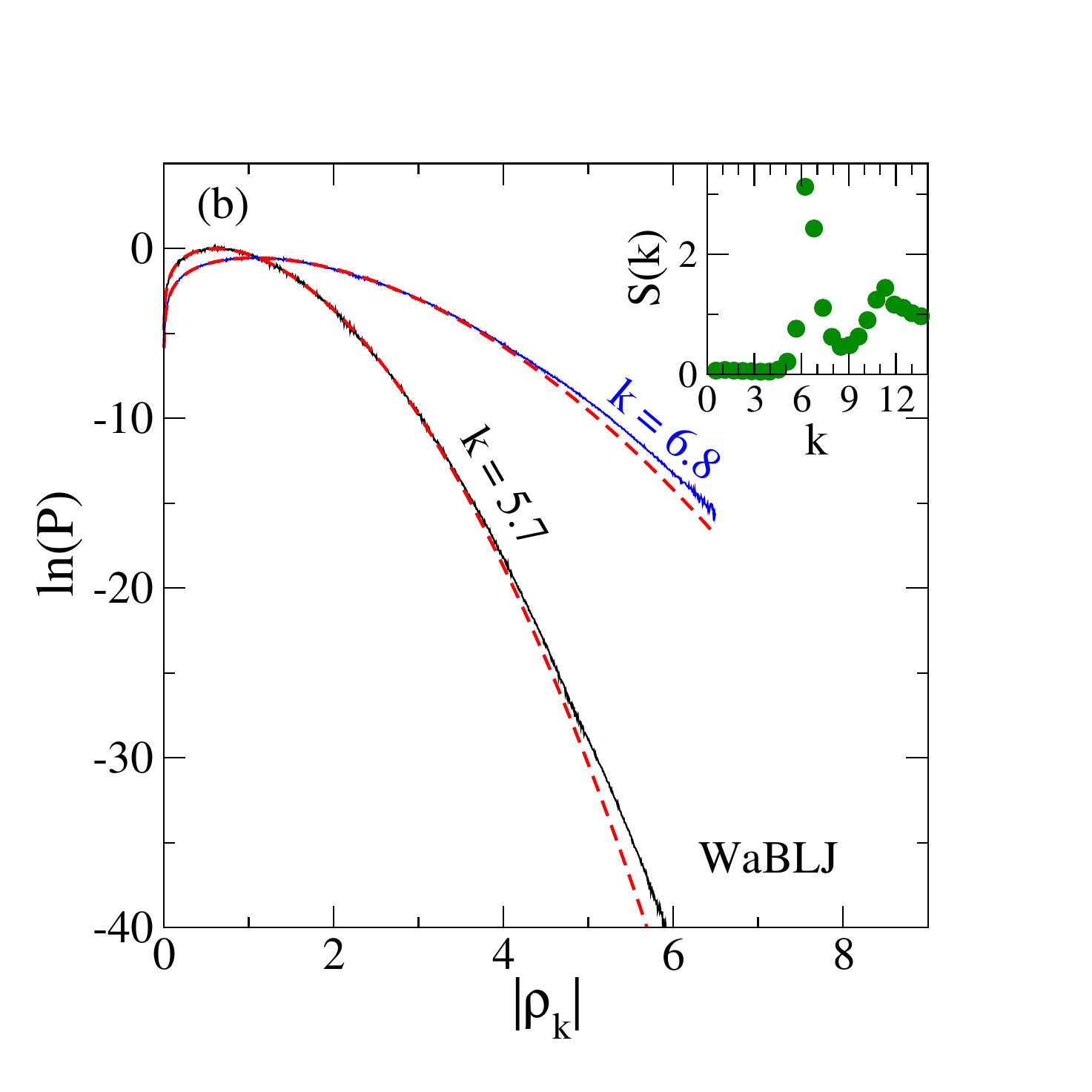}
		\includegraphics[width=0.4\textwidth]{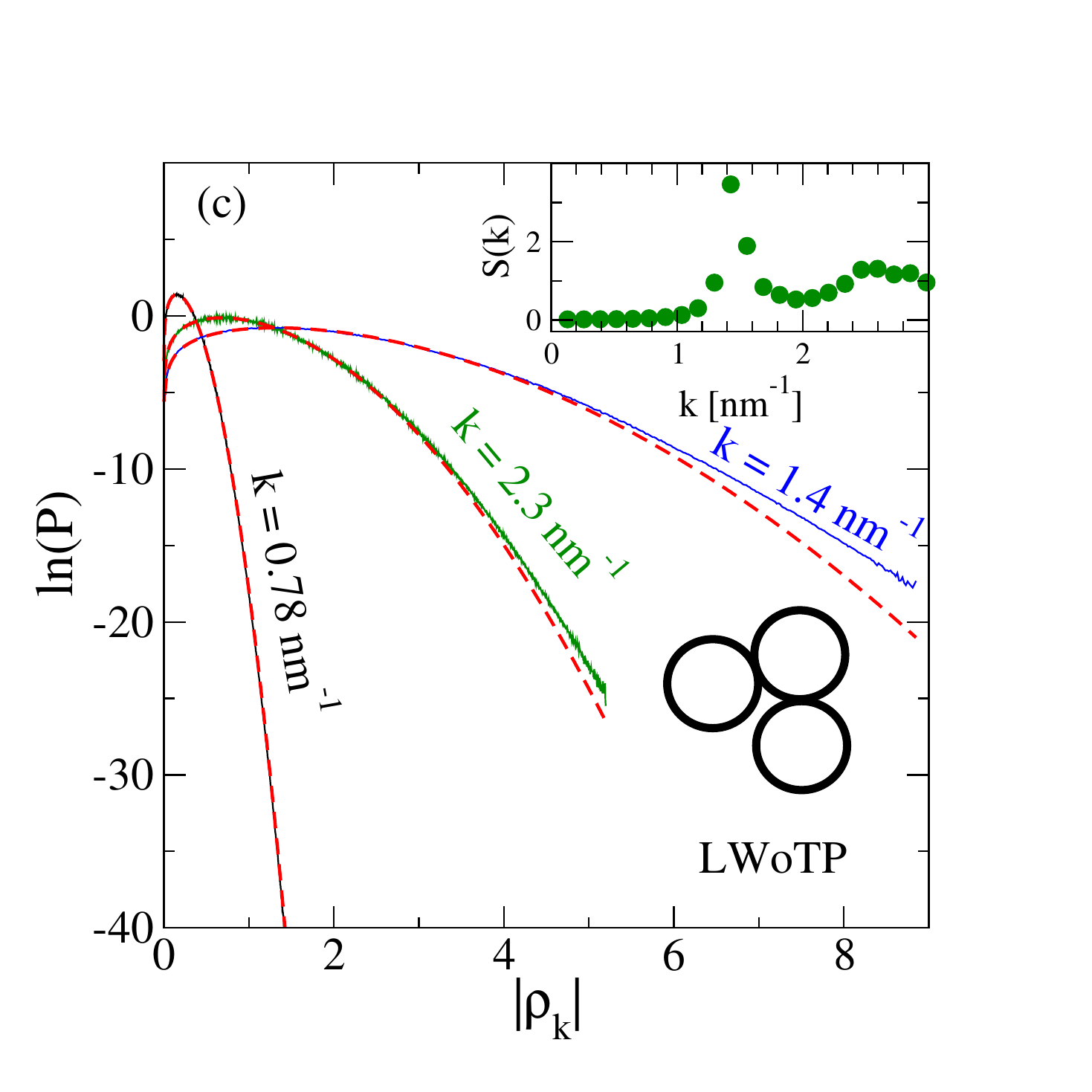}
		\includegraphics[width=0.4\textwidth]{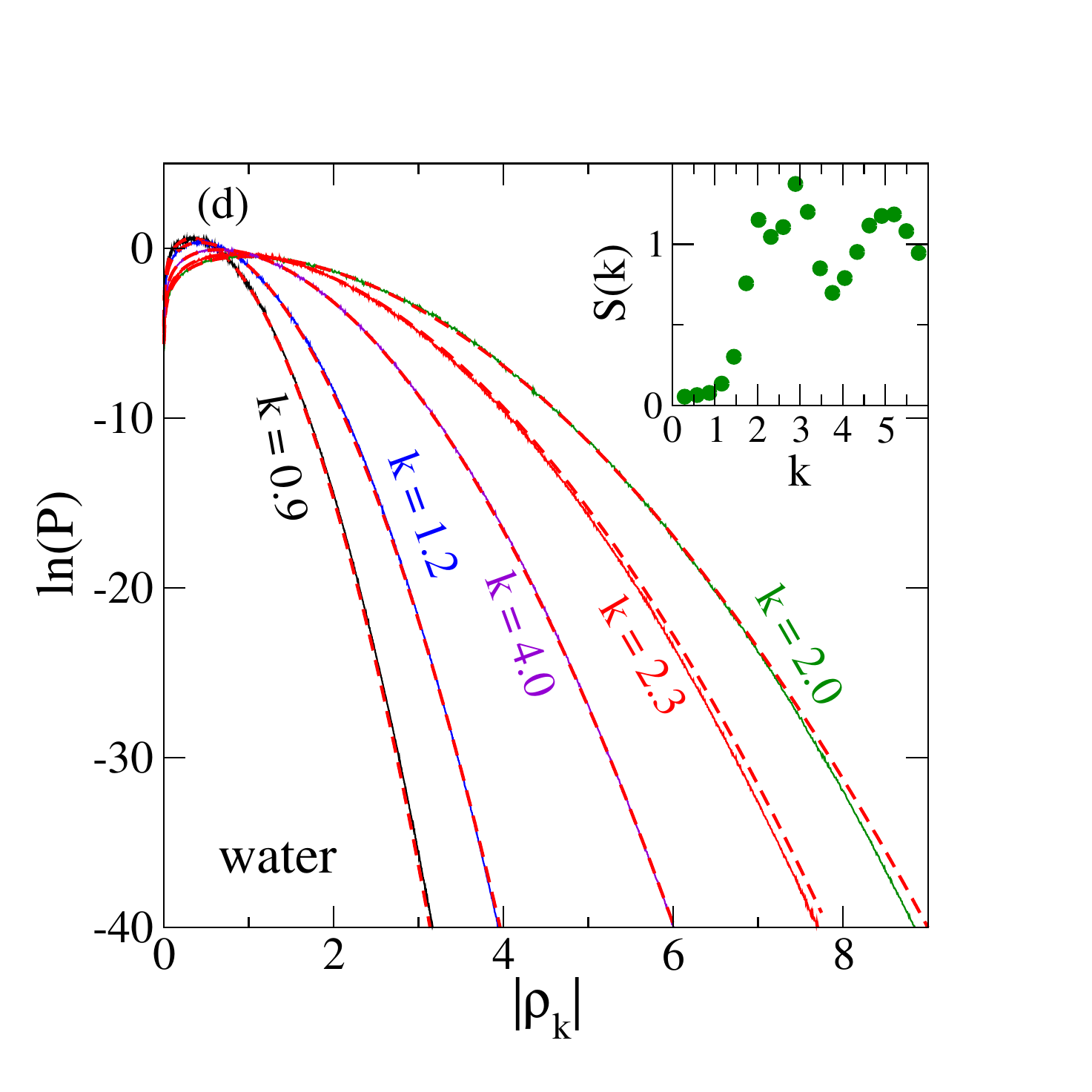}
	\end{center}
	\caption{\label{rhok_stats} Natural logarithm of probability distribution $P(|\rho_k|)$ for (a) the KABLJ mixture ($T=0.45$, $\rho=1.2$, $N=1000$) (b) the WaBLJ mixture ($T=0.6$, $\rho=0.75$, $N=1024$) (c) LWoTP trimer molecules ($T=350$ K, $\rho=1.09$ g/ml, $N=324$), and (d) tip4p/ice water model ($T=280$ K, $\rho=$ 1 g/ml, $N=432$). Red dashed lines are Gaussian predictions. The insets shows scattering functions. The agreement with the Gaussian prediction is good, however, deviations are apparent in the tails of the distributions. These are likely attributed to crystalline structures.}
\end{figure*}

\subsection{Energy surfaces}\label{Models}
We investigate statistics of density fluctuations for several models defined as a $3N$ dimensional energy surface. The examples have been chosen to represent different chemical classes of liquids.
\begin{description}
	\item[LJ] In 1924 Lennard-Jones suggested a simple model of interaction between atoms by summing repulsive term representing Pauli repulsions and an attractive term representing London forces.\cite{lj24} In this study we investigate a truncated version: the potential energy surface is a sum over pair energies $U({\mathbf R})=\sum^N_{n>m} u(|{\bf r}_m-{\bf r}_n|)$ with $u(r)=4\epsilon[(\sigma/r)^{12}-(\sigma/r)^{-6}-2.5^{-12}+2.5^{-6}]$ if $r/\sigma<2.5$ and zero otherwise. The LJ model is not a good glass-former, however, the it is included in this study since it is a standard system in computational condense matter physics. Temperature $T=0.8$ and density $\rho=0.85$ ($L=5.0273$) is used as a representation of the ``normal liquid regime''. This state is close to the freezing temperature.\cite{han69,ped13}
	\item[KABLJ] Kob and Andersen suggested a binary LJ mixture as model of a good glass former.\cite{kob94} It is an 80/20 mixture of particles that have a strong affinity towards unlike atoms. This parametrization makes a good glass-former on time-scales and system sizes typically investigated {\it in silico}. This model is the standard model for computational studies of low temperature liquid dynamics. It is custom to study the system at the density $\rho=1.2$ where the melting temperature is $T_m = 1.027(3)$.\cite{ped18} Below this temperature the particles will eventually phase separate in long-time simulation. The major constituent, the A's, will form a face centered cubic crystal. If crystallization is avoided, however, the low-temperature liquid accumulate locally preferred structures where one of the small particle is surrounded by ten larger particles forming a twisted bicapped square prism \cite{cos07,tur18b}.
	\item[WaBLJ] Wahnst{\"o}m suggested a 50/50 binary LJ mixture with a size ration of 80\% \cite{wah91}. Unlike the KABLJ mixture the interactions parameters ($\varepsilon$'s and $\sigma$'s) follow the Lorentz-Berthelot rule of mixing. The system is a good glass former ({\it in silico}), however, in long-time simulations the mixture can form a MgZn$_2$ crystal structure \cite{ped10}. In the supercooled regime the liquid collect local structures of icosahedral order and Frank-Kasper order.\cite{cos07,ped10,tur18} The latter is a geometric arrangement where two touching larger particles have six smaller particles as common neighbors. These structures are favored by the low-temperature liquid since they pack space well, and are also a part of the crystalline structures.\cite{ped10,kum08}  %Structures like this is a possible candidate for non-Gaussian statistics.
	\item[LWoTP] Lewis and Wahnstr{\"o}m \cite{lew93} suggested a coarse grained model of ortho-terphenyl (C$_{18}$H$_{14}$) where molecules are constructed from three LJ particles placed in the corners of an isosceles triangle. Each LJ particle represent a benzene ring. To avoid that LJ particles crystallize into a close-packed structure the molecule have an inner angle of 75$^\circ$ (that is in between 60$^\circ$ and 90$^\circ$ degree -- the angles found between neighbor triplets in close packed structures). In long-time simulations, however, the system can crystallize into a structure where the LJ particles form a base centered cubic lattice with random orientations of molecule.\cite{ped11a,ped11b} We study a system of $N=324$ molecules (unless else stated) at the temperature $T=350$ K at density $\rho=1.09$ g/ml ($L=4.84$ nm).
	\item [Water] Abascal et al.\ \cite{aba05} suggested the TIP4P/Ice atomistic water model. This four site model reproduce the complicated phase-diagram of real water, suggesting that it also gives a good representation of hydrogen-bonds in the liquid state. The model is studied at temperature $T=280 K$ and density $\rho$ = 1 g/ml. There are no signs of spontaneous crystallization.
\end{description}
Numerical computations are performed using the software packages LAMMPS \cite{lammps}, RUMD \cite{rumd}, and home-made code available at the website \url{http://urp.dk/tools}. Implementation of the $\rho_{\bf k}$ bias field is available in the official LAMMPS and RUMD packages. $\rho_{k}$ statistics are investigated in the constant NVT ensemble. $\rho_h$ fluctuations are studied in systems with a gas-liquid interface. This is done by constructing an elongated simulation cell with periodic boundaries in the $x$ and $y$ directions and walls at the boundaries of the longer $z$ direction.

Results for the LJ, KABLJ and WaBLJ models are reported in reduced Lennard-Jones units, while physical units units are used for the LWoTP model and the TIP4P/ice water model.

\section{Results}\label{Results}

\begin{figure}
	\begin{center}
		\includegraphics[width=0.9\columnwidth]{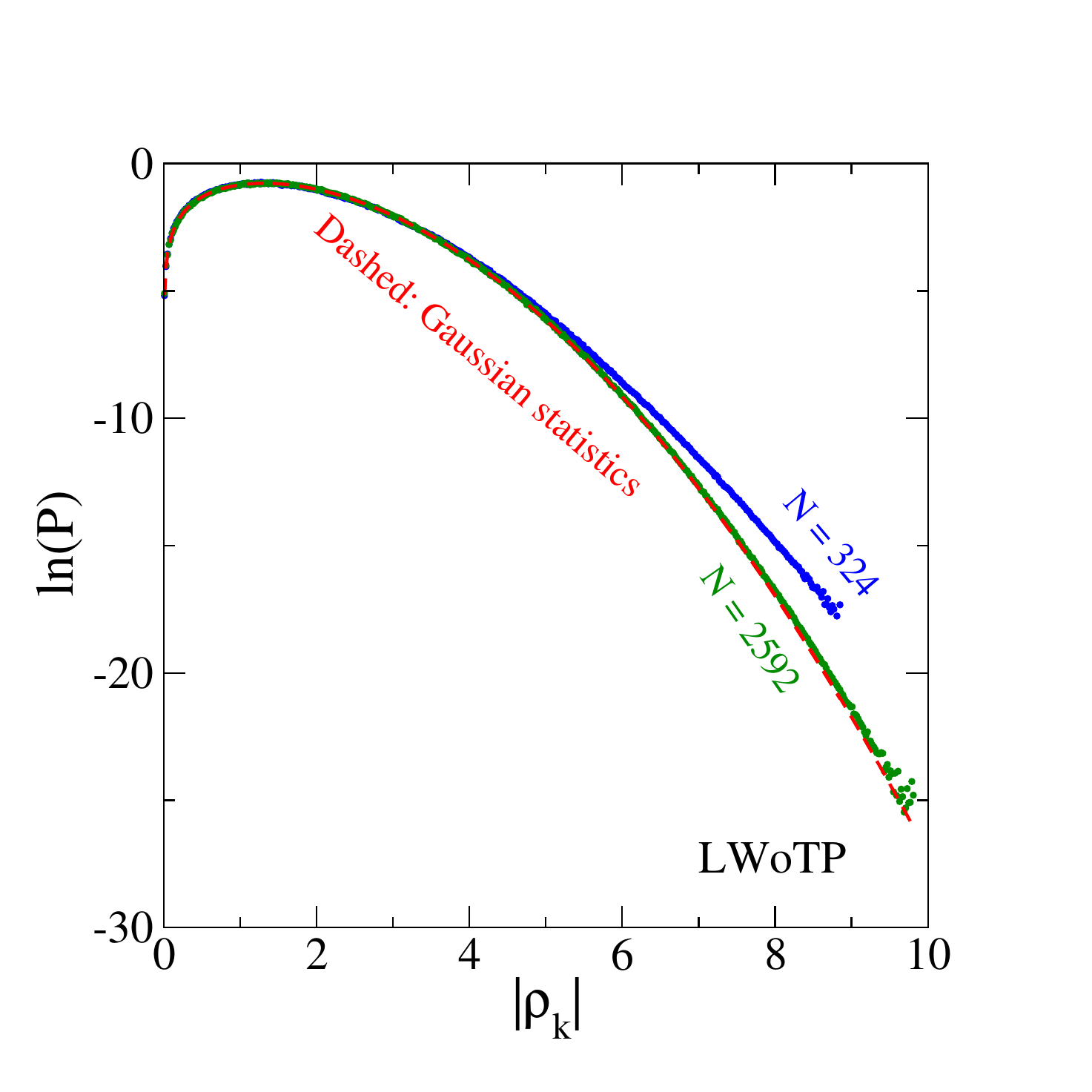}
		\caption{\label{rhok_lwotp_Nscale} The $P(|\rho_{\bf k}|)$ distribution on a logarithmic scale ($k=1.4$ nm$^{-1}$) of the LWoTP model for system sizes of $N=324$ and $N=2592$ molecules, respectively. The Gaussian is better for the larger system as expected from the central limit theorem. The non-Gaussian parameters $\alpha_{\rho_\mathbf{k}}$ are 0.029 and 0.0011 for $N=324$ and $N=2592$, respectively.}
	\end{center}
\end{figure}

\begin{figure}
	\begin{center}
		\includegraphics[width=0.9\columnwidth]{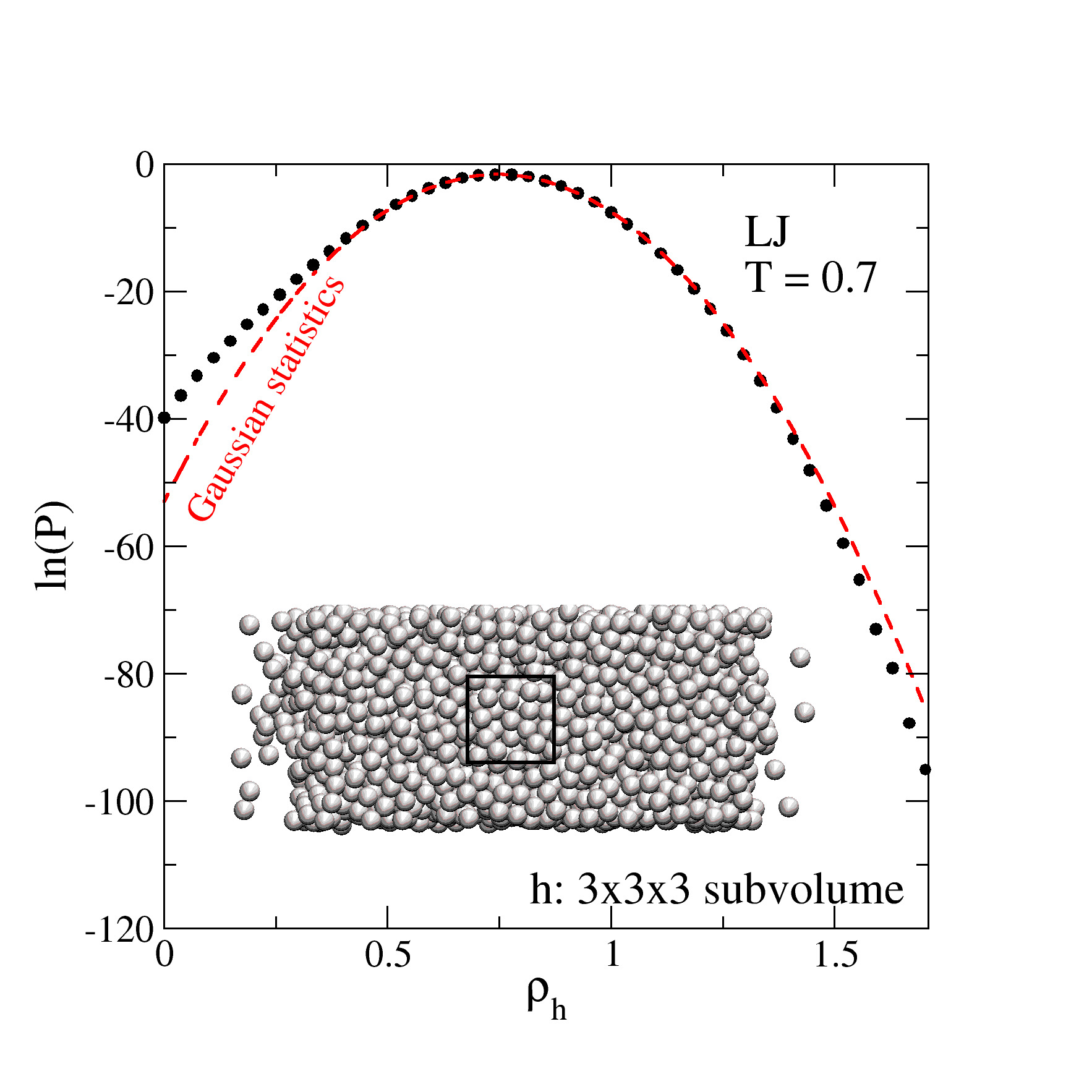}
		\caption{\label{squoplet} Density fluctuations in a $3\times3\times3$ subvolume, $h$, of the LJ model in the normal liquid regime with a gas-liquid interface ($T=0.7$, $N=3000$, $L_x=L_y=10$). The inset show a typical configuration of the system with an gas-liquid interface, and the subvolume $h$ located in the bulk liquid part.}
	\end{center}
\end{figure}

\begin{figure}
	\begin{center}
		\includegraphics[width=0.8\columnwidth]{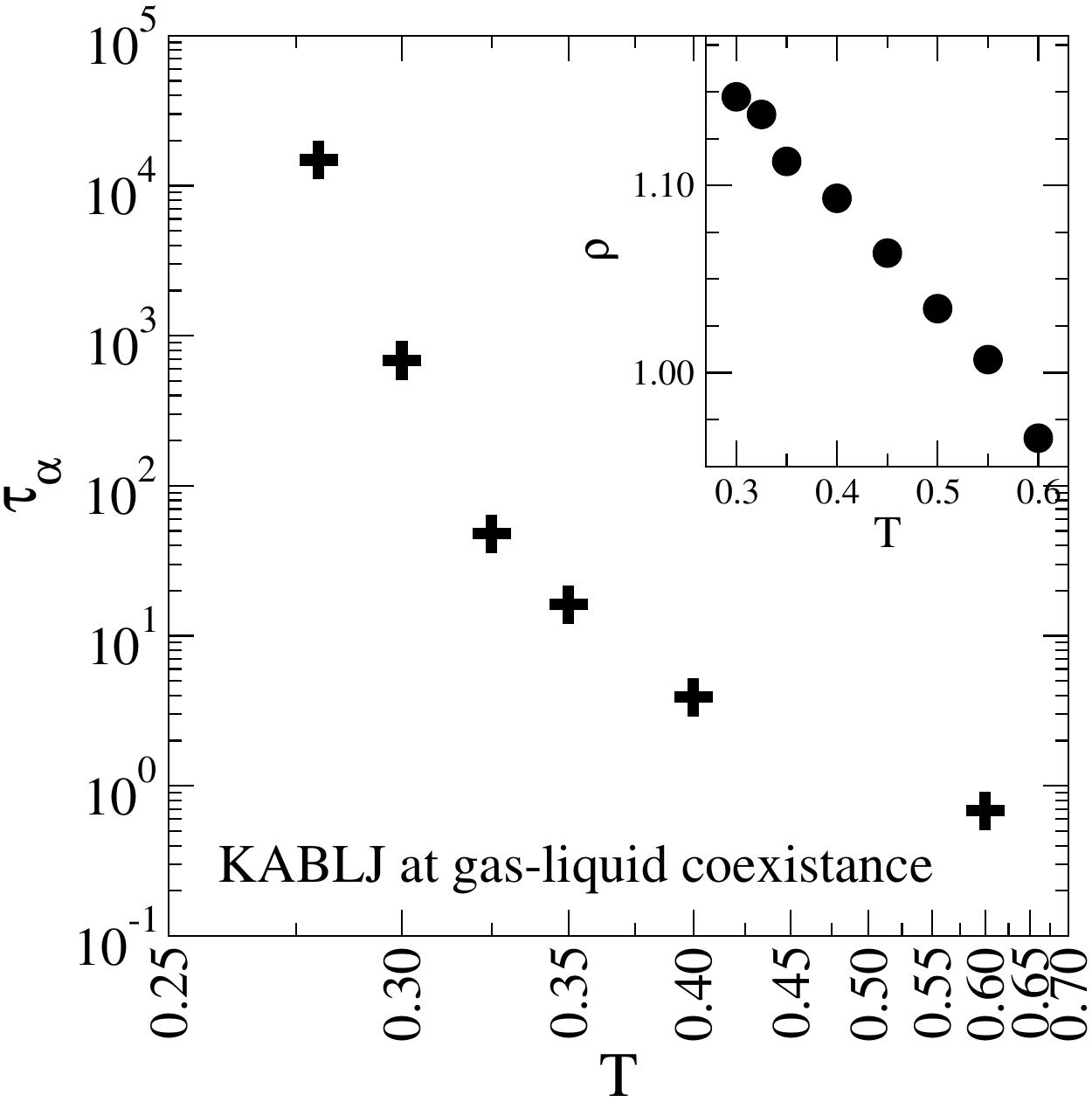}
		\caption{\label{KAslap} Structural relaxation time $\tau_\alpha$ as a function of temperature of KABLJ mixture with gas-liquid interface ($N=3000$, $L_x=L_y=10$). The structural relaxation time is here defined as $F_s(k=2\pi,t=\tau_\alpha)=1/e$ where $F_s$ is the self intermediate scatter function of A's located inside the slap ($-5<z<5$). The inset show the density of the liquid slap as a function of temperature.}
	\end{center}
\end{figure}

\subsection{Fluctuations of the collective density field}

Before investigating the glass forming liquids, we first focus on the LJ model near the melting temperature. Figure \ref{lj108} shows probability distributions $P(|\rho_{\bf k}|)$ on a logarithmic scale. The figure includes $k$-vectors from lengths of $k=1.25$ ($n=1$) up to $k=15.0$ ($n=12$).
The solid black lines are the reweighed distributions using a series of biased simulations and the red dashed lines are the Gaussian predictions (shifted vertically for clarity). The first impression is that the Gaussian hypotheses give a good description. This confirm the consensus that small length scales fluctuations are Gaussian in the normal liquid regime.\cite{hum96,cro97} The tails of the distributions, however, show non-Gaussian features.
As an example, the $k$-vectors with $n=6$ and $n=8$ show fat tails (compared to the Gaussian reference). A representative configuration from the tail of the distribution for $n=6$ is shown on Fig.\ \ref{ljConfs}(a). A crystalline structure is apparent in both the real-space configuration and the scattering spectrum shown on Fig.\ \ref{ljConfs}(b). Consistent with this, a cubic box with 108 LJ particles have an ideal crystal structure with $3\times3\times3$ fcc unit cells giving a Bragg peak at $n=6$. The distribution of the $n=8$ vector (Fig.\ \ref{ljConfs}(c) and Fig.\ \ref{ljConfs}(d)) also have a fat tail. This can also be attributed to a crystalline configuration, but with another orientation. As an aside, bias simulation similar to the ones presented here, can be used to compute the melting point of crystals. This is refereed to as the ``interface pinning'' method.\cite{ped13}

Other $k$-vectors have thin tails. As an example Figs.\ \ref{ljConfs}(e) show a configuration from the tail of the $n=10$ wave vector. This structure it not crystalline but have disordered. Figs.\ \ref{ljConfs}(g) and \ref{ljConfs}(h) show a configuration from the longest wave vector of the investigated system size ($n=1$). The liquid have responded to a strong bias field by forming a vapor slap and a crystalline slap.

Next we investigate the glass forming models. First, we consider the KABLJ mixture at $T=0.45$ ($\rho=1.2$) of a system size of $N=1000$ particles. This state-point is well below the melting temperature of $T_m=1.027(3)$\cite{ped18}. The structural relaxation is about $10^3$ times larger than in the normal liquid regime.\cite{kob94,ped10b} The solid black lines on Fig.\ \ref{KABLJ_umb} shows $P(\rho_{\bf k})$ distributions of the fluctuations of the collective density field for biased simulations using $\kappa=4$ and a range of anchor points. The red dashed lines are Gaussian hypothesis. The agreement is good. 

The systems are more prone to crystallization when a large $|\rho_{\bf k}|$ bias field is applied. Examples of crystallizing trajectories of the KABLJ mixture and LWoTP trimer are shown on Fig.\ \ref{KABLJ_umb}. Crystallizing trajectories are discarded in the analysis.

Figure \ref{rhok_stats} show $P(|\rho_{\bf k}|)$ distributions of the glass forming liquids KABLJ, WaBLJ, LWoTP and water for several $k$-vectors. The red dashed lines indicate the Gaussian approximation. The agreement is good, but the tails of the distributions deviates from the Gaussian prediction. The deviations are system-size dependent as expected from the central limit theorem. Figure \ref{rhok_lwotp_Nscale} shows that the non-Gaussian fat tail for $k=1.4$ nm of the LWoTP systems is greatly diminished when the system size increased from $N=324$ to $N=2592$.

\begin{figure}
	\begin{center}
		\includegraphics[width=0.9\columnwidth]{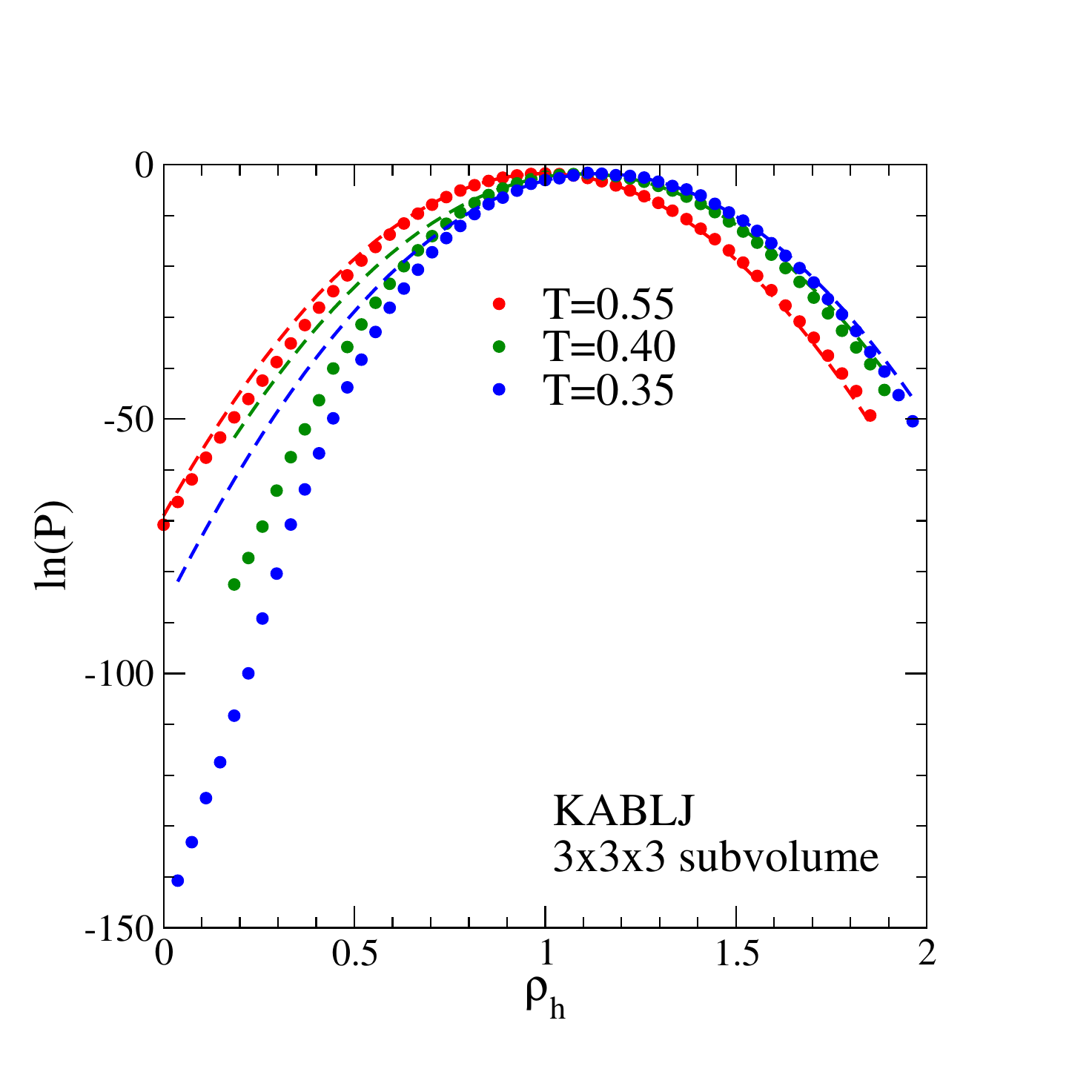}
		\caption{\label{lnP_Nsub_3x3x3} Density fluctuations at different temperatures in a $3\times3\times3$ subvolume of the KABLJ mixture with a gas-liquid interface. The Gaussian approximation becomes better at higher temperature. The non-Gaussian parameter (Eq.\ \ref{nonGauss_h}) for the high temperature $T=0.55$ is $\alpha_{\rho_h}=3\times10^{-4}$ while it is $\alpha_{\rho_h}=-4\times10^{-2}$ for the two lower temperatures, $T={0.40,0.35}$. We contribute low-temperature deviations from Gaussian statistics to subcritical crystallites shown on Fig.\ \ref{lnP_2d}. }
	\end{center}
\end{figure}

\begin{figure}
	\begin{center}
		\includegraphics[width=0.9\columnwidth]{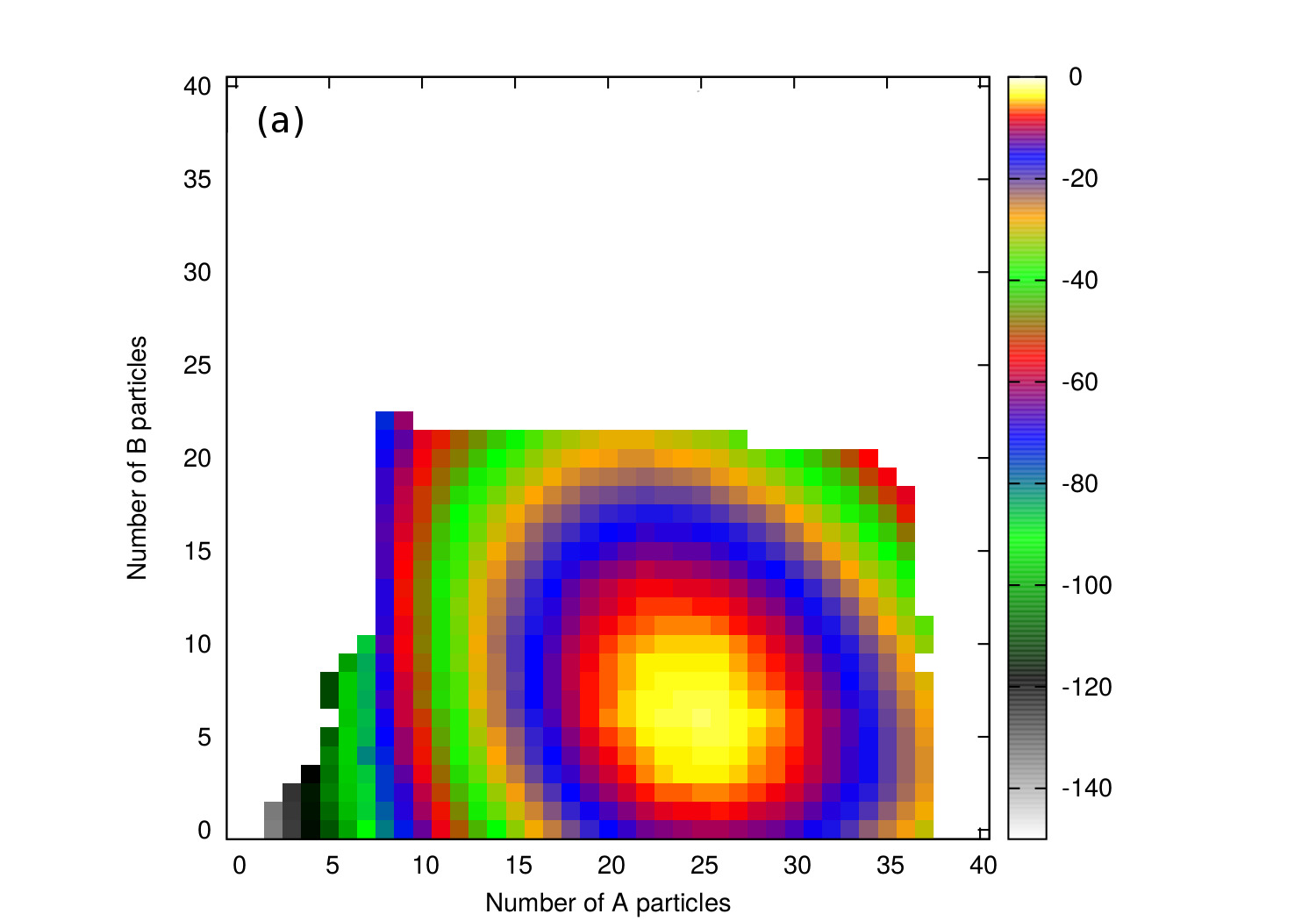}
		\includegraphics[width=0.6\columnwidth]{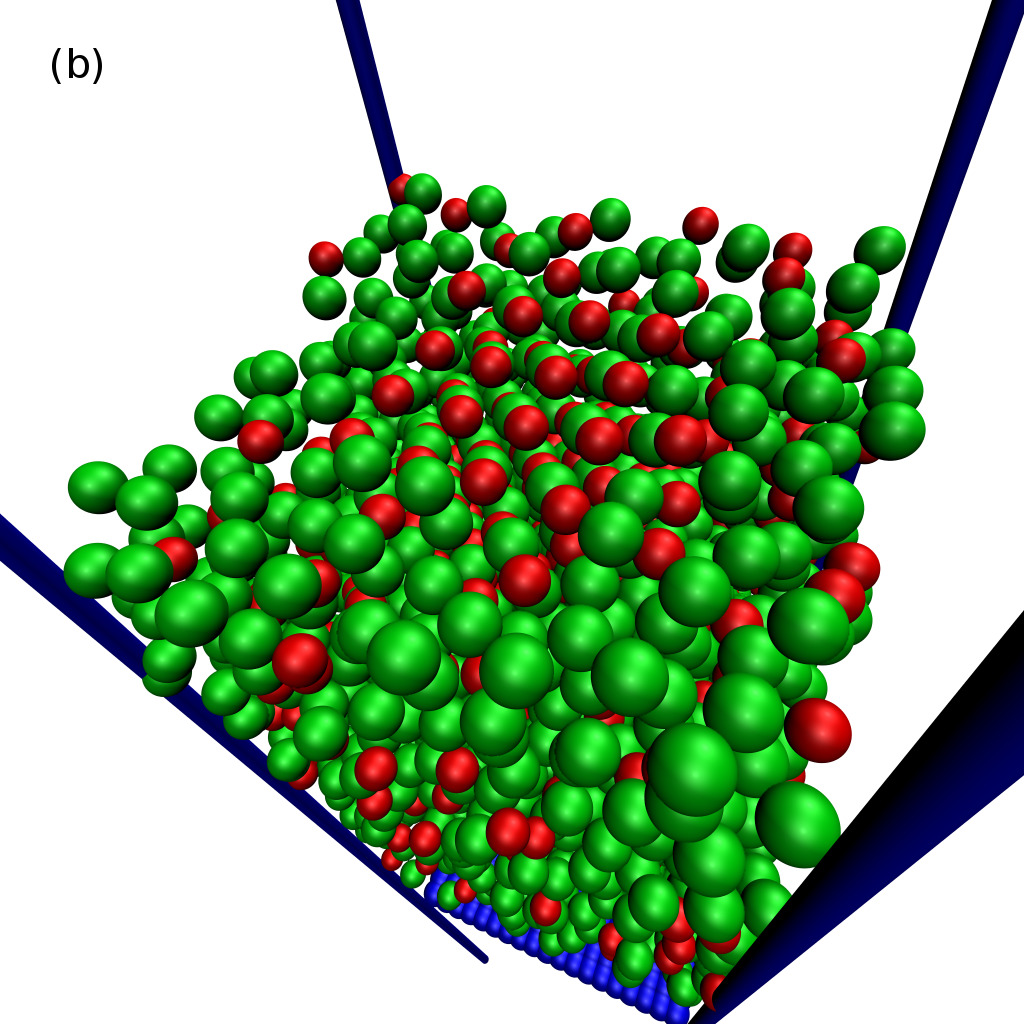}
		\caption{\label{lnP_2d} (a) The $\ln(P(N_A,N_B))$ distribution in a $3\times3\times3$ subvolume of the KABLJ mixture at $T=0.325$ (the white squares were not computed due to bad statistics). Non-gaussian features are seen as the contour lines that deviates slightly from being oval. (b) A configuration from the tail of the distribution with equimolar composition . The upper half of particles have been made invisible to reveal the arrangement of particles in the $3\times3\times3$ subvolume $h$. The structure correspond to a cubic CsCl crystallite. This is one of the known stable structures of the mixture.\cite{ped18}}
	\end{center}
\end{figure}

\subsection{Density fluctuations in a subvolume}
Fluctuations in small subvolumes of a larger system can give a further insight to the structural origin of non-Gaussian small length scale density fluctuations. Figure \ref{squoplet} show the distribution function of the $\rho_h$ density in a $3\times3\times3$ subvolume, $h$. The points are reweighed statistics from simulations with a bias potentials that push the system towards configurations with a certain amount of particles in the subvolume. The red dashed line is the prediction from the Gaussian approximation. The agreement is good,\cite{hum96,cro97} however, some deviations are seen in the tails of the distributions. The low density limit correspond to the formation of a cubic vapor bubble in the liquid.  As described by classical nucleation theory the free energy $-k_BT\ln(P)$ for forming such a bubble is associated with both a bulk- and a surface contribution.

To investigate the supercooled regime we setup a gas-liquid coexistence simulation of the KABLJ mixture (in the same way as we did for the single component LJ model). Figure \ref{KAslap} shows the structural relaxation time in liquid slap. The relaxation time is non-Arrhenius in the investigate temperature regime, $0.28<T<0.60$. The points on Fig.\ \ref{lnP_Nsub_3x3x3} shows the distribution of $\rho_h$ fluctuations in a $3\times3\times3$ subvolume $h$ for the temperatures $T=\{0.35,0.40,0.55\}$. Gaussian statistics are shown as red dashed lines. The conclusion from the analysis of the $|\rho_{\bf k}|$ fluctuations remains -- the Gaussian approximation gives a fair description, but becomes increasingly less accurate at lower temperatures. Deviations are both seen in the tails of the distribution, and near the mean as shown by the non-Gaussian parameter $\alpha_{\rho_h}$. Figure \ref{lnP_2d}(a) shows the $P(N_A,N_B)$ distribution. We would expect to see elliptical shaped contour lines for Gaussian approximation, but see some deviations from this. Figure \ref{lnP_2d}(b) show a configuration from a fat-tail part of the distribution at equimolar composition in the subvolume. The configuration is a cubic CsCl crystallite. This structure is one of the thermodynamically stable crystal structures of the KABLJ model.\cite{ped18} Thus, I conclude that non-Gaussian feature are related to the first order-transition to a crystal.

\section{Discussion}\label{Discussion}
Let us summarize the results before moving on with a discussion of the implications. I have presented an investigation of a range of chemically different glass formers, and the overall conclusion is that the Gaussian hypothesis gives a fair description of the small length scale density fluctuations, however, as temperature is lowered the Gaussian approximation is less accurate.

Gaussian statistics usually comes about in two ways: (i) from the central limit theorem, or (ii) from an harmonic approximation: (i) The central limit theorem dictates that if random variables from any underlying distribution are added the resulting distribution will follow Gaussian statistics. For a non-flowing equilibrium liquid of a sufficient size it can be assumed that subvolumes fluctuate independently. As an example think of the ideal gas model of non interacting particles. For the ideal gas the number of particles in a given subvolume $h$ follow the Poisson distribution. If the size of the subvolume is increased, then Raikov’s theorem dictates that fluctuation follow another Poisson distribution with a higher average. This distribution will be closer to the Gauss function. For a liquid with interactions the underlying distribution differs from Poisson statistics,
however, by studying small length scale density fluctuations we can gain insight on the nontrivial underlying distribution. This brings us the other less trivial way of arriving at Gaussian statistics,
(ii) i.e. by an harmonic expansion around a local minimum of the free energy function $F$: If $x$ is a order parameter, like $\rho_{\bf k}$ or $\rho_h$, then the free energy along this coordinate is 
\begin{equation}\label{free}
F(x)=-k_BT\ln(P(x))
\end{equation}
where $P(x)$ is the probability distribution. The function $F(x)$ can be expressed as a polynomial expansion around the minimum at $x_0$. It is often convenient to assume that only the second order term is of relevance, thus giving a harmonic approximation for the free energy\cite{kar07} 
\begin{equation}\label{freeHarmonic}
F_G(x)\simeq a_2[x-x_0]^2+\textrm{const.}
\end{equation}
The truncation of the expansion series is non-trivial, as discussed below.
By equating Eqs.\ \ref{free} and \ref{freeHarmonic} and isolating $P$ we arrive at Gaussian statistics for the probability distribution
\begin{equation}\label{Pgauss2}
P_G(x) = \exp(-a_2[x-x_0]^2/k_BT).
\end{equation}
(I remind the reader that the reason we find near-Gaussian statistics is {\it not} due to the harmonic bias field).
To understand first order transitions, e.g. the gas-liquid transition, higher order terms are important. As a classic example, Landau's (L) effective Hamiltonian \cite{lan37,tro05} includes higher-order terms to give a description of the density fluctuations near the gas-liquid critical point: 
$F_\textrm{L}(x) = a_2[x-x_0]^2 + a_4[x-x_0]^4 + \textrm{const.}$. 
With the Landau theory in mind, we expect deviations from Gaussian statistics for liquid density fluctuations when other phases (gas or crystals) interfere with the liquid state. In the normal liquid regime deviation from Gaussian statistics can be due to the formation of a vapor bubble as exemplified on Figs.\ \ref{lj108} and \ref{ljConfs}.
In the supercooled regime the crystal basin of the free energy becomes large suggesting that statistics becomes less Gaussian due to the presence of a crystal. The microscopic picture is the formation of subcritical crystallites (Fig.\ \ref{rhok_lwotp_Nscale}). In agreement with this, the systems are more prone to crystallization when a strong field biasing $|\rho_{\bf k}|$ is applied as exemplified on Fig.\ \ref{KABLJ_umbCry}.

I conjecture that it is possible that non-crystalline structures could be important for the statistics of small length scale fluctuations. To address this more refined methods are needed, and I leave this to future studies. Some structural candidates are the ``locally favored structures'' identified for some of the models \cite{tan98,edi00,wid05,shi06,cos07,roy08,ped10,gal10,pas15,tur18,tur18b}. These structures have been suggested as an important component to understand the dynamics of the highly viscous liquid near the glass transition. It would be a valuable insight to show that statistics of small length scale density fluctuations is related to dynamics.

Another angle is to view small length scale density fluctuations from is the ``energy landscape'' perspective \cite{gol69,sas98,vor07,heu08}. In this picture, the $3N$ dimensional energy surface of the liquid is partitioned into basins identified by local minimums. Below a certain onset temperature the system explores confrontational space by two mechanisms. At short times the system vibrates in a basin that is, to a good approximation, harmonic. Thus, it is expected that these vibrations will give rise to Gaussian statistics of density fluctuations. On longer time-scales the system will explore different basing (activated relaxation). From this perspective the non-Gaussian feature at low temperatures are related to density fluctuations between basins (the inherent states). However, this needs to be investigated.

Some theories directly or indirectly assume Gaussian statistics of small length scale density fluctuations.
As an example, the fact that small length scale fluctuations persist to be near-Gaussian in the supercooled regime is in line with the picture behind the generic kinetic contained models.\cite{fre84,cha10,gar02,hed09,rit03} As mentioned in the introduction, the picture is that thermodynamics and structural details of glass forming materials are note crucial for understanding the dynamics of a highly viscous liquid.
Finally, the Gaussian approximation of density fluctuations play a role in some elastic models.\cite{kra40,dyr96,dyr06b} The idea behind these approaches to understand viscous dynamics is though elastic deformation that allow a small length scale subvolume to rearrange. Here the harmonic approximations enters in some theories to make predictions related to experiments.

\section{Acknowledgments}
This study was initiated by discussion with the late David Chandler. I joined his group in 2010-2012, and I greatly benefited from interactions with Patrick Varilly, Amish J. Patel, Thomas Speck, David Limmer, Lester O. Hedges, Yael Elmatad and Adam P. Willard. David had a remarkable talent of challenging conventional beliefs and thereby moving the scientific field forward. For the preparation of this manuscript I also received comments and suggestions from Andreas Tr{\"o}ster, Jeppe C. Dyre and Thomas B. Schr{\o}der.
This work was supported by the VILLUM Foundation’s Matter Grant No. 16515.

% \bibliography{bibliography}


\end{document}